\documentclass[11pt]{article}
\usepackage{jheppub}

\usepackage{epsfig}
\usepackage{amsbsy}
\usepackage{dsfont}
\usepackage{bbm}
\usepackage{upgreek}
\usepackage{amssymb,amscd}
\usepackage{graphicx}
\usepackage{mathrsfs}
\usepackage{amsmath,amsthm}
\makeatletter
\DeclareFontFamily{OMX}{MnSymbolE}{}
\DeclareSymbolFont{MnLargeSymbols}{OMX}{MnSymbolE}{m}{n}
\SetSymbolFont{MnLargeSymbols}{bold}{OMX}{MnSymbolE}{b}{n}
\DeclareFontShape{OMX}{MnSymbolE}{m}{n}{
    <-6>  MnSymbolE5
   <6-7>  MnSymbolE6
   <7-8>  MnSymbolE7
   <8-9>  MnSymbolE8
   <9-10> MnSymbolE9
  <10-12> MnSymbolE10
  <12->   MnSymbolE12
}{}
\DeclareFontShape{OMX}{MnSymbolE}{b}{n}{
    <-6>  MnSymbolE-Bold5
   <6-7>  MnSymbolE-Bold6
   <7-8>  MnSymbolE-Bold7
   <8-9>  MnSymbolE-Bold8
   <9-10> MnSymbolE-Bold9
  <10-12> MnSymbolE-Bold10
  <12->   MnSymbolE-Bold12
}{}

\let\llangle\@undefined
\let\rrangle\@undefined
\DeclareMathDelimiter{\llangle}{\mathopen}%
                     {MnLargeSymbols}{'164}{MnLargeSymbols}{'164}
\DeclareMathDelimiter{\rrangle}{\mathclose}%
                     {MnLargeSymbols}{'171}{MnLargeSymbols}{'171}
\makeatother

\def\be{ \begin{equation} }
\def\ee{ \end{equation}}

\newcommand{\eq}[1]{\begin{align}\begin{split}#1\end{split}\end{align}}

\def\vr{{\vec{r}}}

\def\cot{{\rm cot}}

\def\dim{{\rm dim}}
\def\exp{{\rm exp}}

\def\ker{{\rm ker}}

\def\sign{{\rm sign}}

\def\half{\frac{1}{2}}
\def\thalf{\frac{3}{2}}

\def\ihalf{\frac{i}{2}}

\def\ghalf{\frac{1}{2g^2}}

\def\one{{\hbox{ 1\kern-.8mm l}}}

\def\vx{{\vec{x}}}

\def\vT{{\vec{T}}}
\def\vI{{\vec{I}}}
\def\vJ{{\vec{J}}}
\def\vK{{\vec{K}}}
\def\vS{{\vec{S}}}
\def\vL{{\vec{L}}}
\def\vA{{\vec{A}}}

\def\bz{\bar{z}}

\def\bJ{\bar{J}}

\def\bz{\bar{z}}

\def\hatw{{\hat{w}}}

\def\CD {{\cal D}}

\def\CJ {{\cal J}}
\def\CK {{\cal K}}
\def\CL {{\cal L}}

\def\CN {{\cal N}}

\def\CR {{\cal R}}

\def\IC{\mathbb{C}}

\def\IR{{\mathbb{R}}}

\def\IZ{{\mathbb{Z}}}

\def\fg{\mathfrak{g}}

\def\fs{\mathfrak{s}}

\def\fsu{\mathfrak{su}}

\def\fu{\mathfrak{u}}

\def\fM{\mathfrak{M}}

\def\cnj#1{\bigskip\noindent{\bf Conjecture:} }

\def\tildeG{{\widetilde{G}}}
\def\tildeH{{\widetilde{H}}}

\DeclareMathAlphabet{\mathpzc}{OT1}{pzc}{m}{it}

\def\Tr{ \, \textrm{Tr} \, }

\def\csch{ \, \textrm{csch} \, }

\def\diag{{{\rm diag\,}}}

\def\vx{{\vec{x}}}

\title{Callan-Rubakov Effect and Higher Charge Monopoles
}

\author[a]{T.~Daniel Brennan}
\affiliation[a]{Kadanoff Center for Theoretical Physics \& Enrico Fermi Institute, \\
University of Chicago,
Michelson Center for Physics, 933 E 56th St, Chicago, IL 60637. 
}

\emailAdd{tdbrennan@uchicago.edu}

\abstract{ In this paper we study the interaction between magnetic monopoles and massless fermions. In the low energy limit, the monopole's magnetic field polarizes the fermions into purely in-going and out-going modes. Consistency requires that the UV fermion-monopole interaction leads to non-trivial IR boundary conditions that relate the in-going to out-going modes. These non-trivial boundary conditions lead to what is known as the Callan-Rubakov effect. 
Here we derive the effective boundary condition by explicitly integrating out the UV degrees of freedom for the general class of spherically symmetric $SU(N)$ monopoles coupled to massless fermions of arbitrary representation. We then show that the boundary conditions preserve symmetries without  ABJ-type anomalies. % but not those with `t Hooft anomalies. 
 As an application we explicitly derive the boundary conditions for the stable, spherically symmetric monopoles associated to the $SU(5)$ Georgi-Glashow model and comment on the relation to baryon number violation. 
}

\begin{document}

 \maketitle

\section{Introduction and Summary}

In the 1980's Callan and Rubakov showed that the interaction of massless fermions with a minimal $SU(N)$ monopole is non-trivial in the IR \cite{Callan:1982ac,Callan:1982ah,Callan:1982au,Rubakov:1982fp,Callan:1983ed,Callan:1983tm}. 
%
%Often when quantizing a theory in a smooth background field configuration, one expects that the IR dynamics are governed by the asymptotic form of the field configuration. However, for the monopole-fermion system, 
%
At low energies, the smooth $SU(N)$ monopole is described by a singular $U(1)$ monopole. % in which the strength of the magnetic field becomes the order of the UV cutoff scale. 
%in the low energy limit which in general may require additional boundary conditions that encode the UV interaction in the monopole core. 
%
%
%
One then expects that the strong magnetic field near the core of the monopole repels dynamical fields and screens the UV physics in the monopole core. However, in this theory the fermions have low-energy spin-$j=0$ modes that  penetrate into the core of the monopole. This allows the fermions to interact with the non-abelian degrees of freedom that are confined to the monopole core and induces a non-trivial effective boundary condition on the IR abelian monopole \cite{Callan:1982ac,Callan:1982ah,Callan:1982au,Rubakov:1982fp,Callan:1983ed,Callan:1983tm,Grossman:1983yf,Yamagishi:1982wp,Yamagishi:1983ua,Yamagishi:1984zu,Panagopoulos:1984ws,Sen:1984qe,Sen:1984kf,Balachandran:1983sw,Goldstein:1983bk,Isler:1987xn}. 

Traditionally, this non-trivial boundary condition is stated in terms of scattering processes where an s-wave fermion scatters off of the monopole. This viewpoint has lead to much confusion in the literature due to the fact that the monopole boundary condition is not stated simply in terms of the fermion fields. The main confusion is how to understand the physical interpretation of the out-state of a typical scattering process \cite{Preskill:1984gd,Kitano:2021pwt} which has been cryptically referred to as ``propagating pulses of vacuum polarization'' \cite{Polchinski:1984uw}. 

% which has been referred to as the ``final state puzzle'' or the ``semiton puzzle'' \cite{Preskill:1984gd,Kitano:2021pwt}. The crux of the issue is the physical interpretation of the out-state of a typical scattering process  
%%In particular, the classic paper of Polchinski \cite{Polchinski:1984uw} models the scattering process as fermions interacting with a ``rotor'' (i.e. charged, periodic scalar field). 
%which has been cryptically referred to as ``propagating pulses of vacuum polarization'' \cite{Polchinski:1984uw}. 

The reason why this fermion-monopole interaction is of great importance is because of its phenomenological implication in the standard model. 
Originally, Callan and Rubakov studied monopoles in the $SU(5)$ Georgi-Glashow GUT completion of the standard model. There, Callan and Rubakov showed that the fermion-monopole interaction violates baryon-number symmetry in a way that catalyzes proton decay. This effect, which is now referred to as the Callan-Rubakov effect, is clearly important in understanding the phenomenological implications of GUT completions of the standard model. 

\subsection{Outline and Summary of Results}

Thus far, the Callan-Rubakov effect has only been discussed for theories in which the fermion-monopole interactions reduce to the spherically symmetric $SU(2)$ monopole coupled to fundamental fermions. 
In this paper we revisit the Callan-Rubakov effect in order to clarify some of the subtle issues described above and we determine the form of the IR boundary conditions for massless fermions of general representation on a generic spherically symmetric $SU(N)$ monopole.

First, in Section \ref{sec:2} we discuss the low energy effective theory of massless fermions in the presence of a  spherically symmetric monopole in $SU(N)$ gauge theory. In the IR, the smooth UV monopole is described by a singular Dirac monopole. In the Dirac monopole background, the magnetic field  
polarizes the fermions so that each IR fermion is of definite angular momentum and purely in-going or out-going. These correspond to the fermions in the lowest Landau level when restricted to a spherical shell surrounding the monopole: the scale of the Zeeman splitting is set by the $W$-boson mass $m_W$ which is on the order of the UV-cutoff scale. Because of the induced energy splitting, the IR theory is well described by the spherical reduction to a 2D theory on the half-plane which describes the radiald propagation of the above modes. 

In the effective 2D theory the in-going/out-going fermion modes correspond to 2D chiral/anti-chiral fermions and the monopole defect corresponds to the boundary on the half-plane at $r=0$. Unitarity then demands that we impose    boundary conditions on the monopole that relate the in-going/chiral to out-going/anti-chiral modes. 
The general boundary conditions for such a system are given by linear relations between the currents of the chiral $(J_A$) and anti-chiral fermions $(\bJ_A)$ which can be written as 
\cite{Smith:2019jnh,Smith:2020rru,Smith:2020nuf}:
\eq{
J_A=\CR_{AB}\bJ_B\big{|}_{r=0}~.
}
The Callan-Rubakov effect corresponds to a particular choice of matrix $\CR_{AB}$, which can be derived directly from integrating out UV degrees of freedom \cite{Callan:1982ac,Dawson:1983cm}. This effective boundary condition arises from the interaction of the fermions with a collection of periodic scalar fields (called dyon degrees of freedom) that are localized on the monopole. 

In Section \ref{sec:3}, we derive the origin of a spherically symmetric monopole's dyon degrees of freedom from the UV theory. % of smooth monopoles. 
Spherically symmetric monopoles are smooth, classical  Yang-Mills-Higgs field configurations that source a magnetic field. 
They are specified by a pair of embeddings $SU(2)_{T,I}\hookrightarrow SU(N)$ that are generated by the triplets $\vT,\vI\in \fsu(N)$. The pair source the asymptotic magnetic field 
\eq{
\lim_{r\to \infty}\vec{B}=\frac{\gamma_m}{2r^2}\hat{r}+O\big(1/r^{2+\delta}\big)\quad, \quad 
\gamma_m=2(T_3-I_3)~,
}
where $\delta>0$ and $[\vI,\gamma_m]=0$. The non-trivial Higgs profile asymptotically breaks gauge symmetry and creates a potential well at the monopole that traps the non-abelian gauge field in the broken directions (referred to as $W$-bosons). This gives rise to a light degree of freedom $\varphi$ which is localized on  the monopole world volume which are the  phase modes of the trapped $W$-bosons. 

\begin{figure}
\begin{center}
\includegraphics[scale=0.8,clip,trim=1cm 22cm 2cm 1cm]{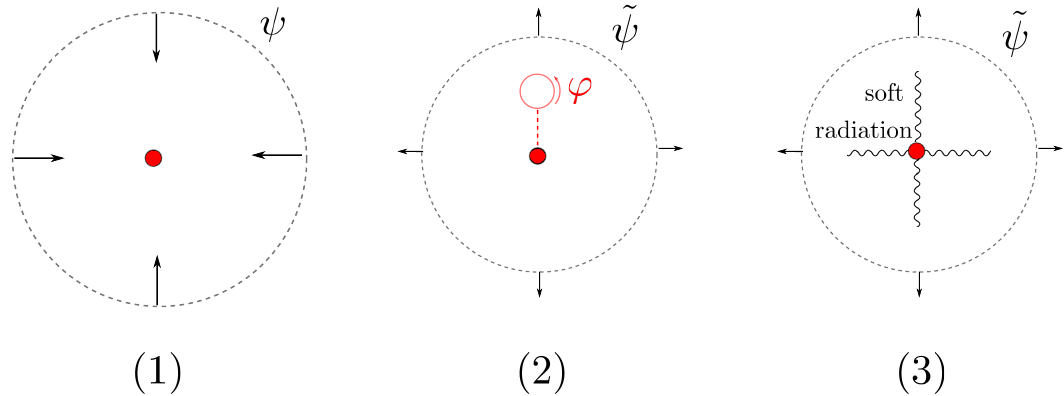}
\end{center}
\caption{In this figure we illustrate the typical scattering of a low energy fermion mode off of a monopole. (1) shows an incoming spherical fermion wave of $\psi$ which is (2) reflected to an out-going spherical mode of a different fermion $\tilde\psi$ while exciting the periodic dyon degrees of freedom $\varphi$. The dyon excitation then (3) decays by soft radiation of the massless, charged fermion modes.}
\label{fig:illustration}
\end{figure}

The dyon degrees of freedom couple to the low energy fermion degrees of freedom through the coupling to the UV gauge field which can be read off explicitly from the spectrum of exact fermion zero-modes. 
%It then remains to derive the coupling of the dyon degrees of freedom to the fermion degrees of freedom in the low energy effective theory. One can read off these couplings explicitly by solving for the exact spectrum of fermion zero-modes. 
With these solutions in hand, one can then integrate out the dyon degrees of freedom to obtain the effective boundary condition on the long range fermion degrees of freedom.

The resulting boundary conditions are qualitatively similar to the boundary conditions found by Callan and Rubakov: for any in-going particle, there is an out-going fractional flux in multiple fermion channels. This boundary condition encodes the following physical process. When a low energy fermion scatters off of the monopole, it excites the dyon degrees of freedom $\varphi$. 
 The scattering deposits charge and energy in the monopole core which is then radiated away by soft, charged modes of the massless fermions:
 \eq{\label{firsteq}
\psi\longrightarrow \tilde\psi+\varphi\longrightarrow \tilde\psi+J_{elec}=\tilde\psi+\frac{1}{\CN}\sum_ic_i\tilde\psi_i^\dagger\tilde\psi_i~,
}
where $\CN$ is some normalization constant. 
This is illustrated in Figure \ref{fig:illustration} and is the content of Section \ref{sec:4}. 
 
The general boundary condition for a collection of fermions $\{\psi_{R_i}\}$ that each transform under representations $R_i$ of $SU(N)$ coupled to a spherically symmetric monopole is given as follows. In total, the number of low energy fermion modes is given by\footnote{
Here $\langle~,~\rangle$ is the natural pairing $\langle~,~\rangle:\fg\times \fg^\vee\to \IR$.}
 \cite{Brennan:2021ucy}
\eq{
\{\#\text{ of low-energy modes}\}=\sum_{R_i}\sum_{\mu\in S_{R_i}}n_{R_i}(\mu)\langle T_3,\mu\rangle~,}
where 
$
S_R=\big\{\mu \in \Delta_R\,\big{|}\,\langle \gamma_m,\mu\rangle\times\langle \gamma_m,\mu^\ast\rangle<0\big\}$ for $\Delta_R$ the weight system of $R$,  
$n_R(\mu)$ is the multiplicity of the weight $\mu\in \Delta_R$, and $\mu^\ast$ is the charge conjugate with respect to the $SU(2)_T$ embedding\footnote{Here the charge conjugate of $\mu$ is the image of $\mu$ under the generator of the Weyl group of $SU(2)_T\subset SU(N)$. 
}. If we label all low energy modes by a general index $A$, then we can define the matrix 
\eq{
c_A^I=\langle \rho_I,\mu_A\rangle\quad, \quad \rho_I=\sum_{J =1}^{I}J( H_J+H_{N-J})+I\sum_{J=I+1}^{N-I-1}H_J~,
}
where $\mu_A$ is the weight of the fermion mode labeled by $A$ and $\{H_I\}$ are simple coroots.  
The effective boundary condition is then specified by the matrix 
\eq{
\CR_{AB}=\delta_{AB}-\sum_{I=1}^{N-1}\frac{2}{\sum_{C}c^I_{C}\langle H_{N-I},\mu_C\rangle}c^I_Ac^I_B~,
}
which is  very non-diagonal and leads to fractional fluxes in a generic scattering process. See Section \ref{sec:4} for more precise definitions and details.

The existence of fractional flux in a given fermion channel comes from the fact that each dyon couples to a corresponding $U(1)$ gauge current whose unit normalization leads to fractional coefficients as in \eqref{firsteq}. 
This peculiarity arises from the fact that we have massless charged fermions which each have an anomalous species number symmetry. This means that the species number is not a good quantum number and has no right to be preserved by a scattering process with monopoles which are non-perturbative gauge field configuration. 
However, the preserved gauge and global symmetries do have good quantum numbers and do have integer fluxes in a generic scattering process. In fact, we show here that the effective monopole boundary conditions explicitly violate global symmetries that are only broken by ABJ-type anomalies.  This is the topic of Section \ref{sec:5}. 

Finally, in Section \ref{sec:6} we demonstrate our results by explicitly computing the effective boundary conditions for a number of examples.  As an application we solve for the effective boundary conditions for the higher charge spherically symmetric monopoles   in the $SU(5)$ Georgi-Glashow GUT model $SU(5)$ and show that they preserve $B$-$L$ symmetry while  breaking $B$- and $L$-symmetry. 

We expect that this work has an interesting extension to the case of massive fermions. In the massive case, the fermion spectra is qualitatively different: there exists an infinite tower of angular momentum modes that are $L^2$-normalizable/bound to the monopole. Additionally, when the fermions have a mass there is a gap to the continuum of scattering states that prevents the dyon degree of freedom from radiating away its electric charge. We hope to study this scenario in a future work.

\section{Low Energy Monopole-Fermion System and the Callan-Rubakov Effect}

\label{sec:2}

We are interested in understanding the low energy monopole-fermion interactions in UV complete non-abelian gauge theories. In this section we will introduce the Callan-Rubakov effect from the low energy effective theory point of view. % and set up the low energy scattering problem. 

 Let us consider a 4D $SU(N)$ gauge theory with an adjoint-valued scalar field $\Phi$ and a fermion $\psi_R$ of representation $R$. Here we will consider the vacuum in which 
 $\Phi$ has a vev $\Phi_\infty$  that breaks 
 \eq{
 SU(N)\longrightarrow G_{\Phi_\infty}=\frac{\tildeG_{\Phi_\infty}}{\Gamma}=\frac{SU(N_1)\times... \times SU(N_n)\times U(1)^r}{\Gamma}~,
 }
   where $\Gamma$ is a discrete abelian subgroup of the center of $\tildeG_{\Phi_\infty}$. Here we will \emph{not }restrict ourselves to the case where $G_{\Phi_\infty}$ is abelian.

% \emph{
%However, the 
%low energy theory is still coupled to a generically non-abelian dynamical gauge field. While we will work in the regime where the weak coupling limit where the theory  is well described by UV fermion degrees of freedom, it is simpler to understand the monopole-fermion interaction in terms of boundary conditions rather than in terms of scattering processes because it allows us to ignore the dynamics of the non-abelian gauge field. }

 This theory has a collection of monopoles which are smooth field configurations that solve the Bogomolny equation %\cite{ref}
\eq{
B_a=D_a\Phi\quad, \quad E_a=0~,
}
where $B_a,E_a$ are the non-abelian magnetic and electric field. Monopole solutions are classified by their asymptotic behavior 
\eq{
\lim_{r\to \infty}B_a=\frac{\gamma_m}{2r^2}\hat{r}+O(1/r^3)\qquad \lim_{r\to \infty}\Phi=\Phi_\infty-\frac{\gamma_m}{2r}+O(1/r^2)~,
}
where $\gamma_m$ is the magnetic charge and $\Phi_\infty$ is the Higgs vev. Here we take  $ \gamma_m=\sum_I n_I h^I$ for $n_I\in \IZ_{\geq0}$ where $\{h^I\}$ are generators of $U(1)^r$ which we take to be normalized $e^{2\pi i h^I}=1$. %Without loss of generality, we will restrict to field configurations with 

Since this theory is in general non-abelian, the theory has two natural scales associated to it: the UV cutoff scale (set by the Higgs vev) and the confinement scale. Here we will consider the ``low energy limit'' at intermediate energy scales where the theory is well described by deconfined fermions coupled to  the $G_{\Phi_\infty}$ gauge fields.

 In this limit, the smooth monopole is described by a monopole defect operator which sources a gauge field:
\eq{\label{abelianconn}
A_{IR}=\gamma_m A_{Dirac}=\frac{\gamma_m}{2}(1-\cos\theta)d\phi~,% \quad \gamma_m=\sum_I n_I h^I~~,~~n_I \in \IZ_{\geq0}~.
}
in spherical coordinates centered around the monopole\footnote{ Without loss of generality we will be restricting ourselves to the northern hemisphere.}. Since this field configuration is singular, the monopole operator is defined by cutting out an infinitesimal 2-sphere around the origin and on which we impose the boundary condition \eqref{abelianconn} on the $G_{\Phi_\infty}$ fields. In general, this procedure requires boundary conditions for any fields whose long distance behavior does not vanish near the cutoff surface. These boundary conditions will encode the UV physics of a dynamical field's interaction with the UV magnetic monopole.

The magnetic charge $\gamma_m$ generates a subgroup $U(1)_m\subset G_{\Phi_\infty}$. The representation $R$ of $SU(N)$ then decomposes into $U(1)_m$ representations as 
\eq{
R=\bigoplus_i R_i\quad, \quad R_i[\gamma_m ]=p_i\in \IZ~.
}
%where each $R_i$ is a representation of $\prod_a SU(N_a)$. We would like to further pick a basis for each $R_A$ so that 
Here we will find it useful to introduce the notion of representation weights. For a representation $R$ there are a collection of weights $\mu\in \Delta_R$. To each weight $\mu$ there is a vector $v_\mu$ in the representation space of $R$ and the set of $v_\mu$ form a basis of the representation space. Every weight $\mu$ can be expressed as a vector in $\IR^{N-1}$ and elements of the Cartan subgroup (spanned by simple coroots $\{H_I\}$) have corresponding (dual) vectors so that the charge of $v_\mu$ with respect to $H_I$ is given by the Euclidean inner product in $\IR^{N-1}$:
\eq{
R(H_I)\cdot v_\mu=\langle H_I,\mu\rangle \,v_\mu~.}
%Given a choice of Cartan subgroup $T\subset G$, the weights of a representation $\mu$ form a collection $\Delta_R$ of integral vectors in $\IR^{r}$ where $r={\rm rank}[G]$. The generators of $T$, which we denote $\{H_I\}$, correspond to (dual) vectors in $\IR^{N-1}$ so that the charge of vector $v_\mu$ with respect to $H_I$ is given by the Euclidean inner product in $\IR^{N-1}$:
%\eq{
%R(H_I)\cdot v_\mu=\langle H_I,\mu\rangle \,v_\mu~.}
Consequently, given a choice of Cartan subgroup containing $U(1)_m$, we can decompose 
\eq{
R=\bigoplus_{\mu_i\in \Delta_R} R_i~,%,\quad p_i=\langle \gamma_m,\mu_i\rangle~~,~~ \mu_i\in \Delta_{R_i}~,
}
for weights $\mu_i\in \Delta_{R}$ in the weight space  of $R$ where $R_{\mu_i}$ has charge $p_i=\langle \gamma_m,\mu_i\rangle$ under $U(1)_m$. Consequently $\psi_R$ also decomposes
\eq{
\psi_R=\sum_{\mu_i\in \Delta_R} \psi_{\mu_i}v_{\mu_i}~.
}
% In other words, $\psi_R$ decomposes into fermions $\{\psi_{\mu_i}\}$ with charges $p_i=\langle \mu_i,\gamma_m\rangle$ under $U(1)_m$. 
Here we will assume that 
%in addition to $\sum_i p_i^3=0$ that the theory has $\sum_i p_i=0$. This comes from assuming that 
the theory has no gauge anomaly. This implies that $\sum_i p_i^3=0$ and $\sum_ip_i=0$%which is the requirement that the theory can e UV completed by a non-anomalous higher rank gauge group
\footnote{
The fact that $\sum_i p_i=0$ follows from the fact that all $U(1)$ generators in $SU(N)$ are traceless. 
}.

In the monopole background, the magnetic field polarizes the fermion spin by the Zeeman effect. This can be seen by computing the spectrum of the Dirac operator. As in Appendix \ref{app:A}, the solutions to the Dirac equation are a spectrum of scattering states built on the zero-mode solutions:
\eq{\label{abeliansolutions}
\psi_{\mu_i} (r,\theta,\phi)=&c_{+,m}\,\frac{e^{i\left(m- \frac{p_i}{2}\right)\phi}}{r} r^{\sqrt{(j+1/2)^2-p_i^2/4}}\,U(\theta,\phi)\left(\begin{array}{c}
 d^j_{-m,\frac{p_i+1}{2}}(\theta)\\
 d^j_{-m,\frac{p_i-1}{2}}(\theta)
\end{array}\right)\\
&+c_{-,m}\, \frac{e^{i\left(m- \frac{p_i}{2}\right)\phi}}{r} r^{-\sqrt{(j+1/2)^2-p_i^2/4}}\,U(\theta,\phi)\left(\begin{array}{c}
 d^j_{-m,\frac{p_i+1}{2}}(\theta)\\
- d^j_{-m,\frac{p_i-1}{2}}(\theta)
\end{array}\right)~.
}
Requiring that the fermion's energy density be finite at $r\to 0,\infty$ then implies that there are no solutions for $p_i=0$ and restricts us to the solutions with 
\eq{
j=j_i:=\frac{|p_i|-1}{2}~,
}
for $p_i\neq 0$. Explicitly, we find that the finite energy density solutions of the Dirac operator are spanned by 
\eq{
\psi_{\mu_i}^{(k,m)}=\frac{U(\theta,\phi) }{r}\hat\psi_{\mu_i}^{(k,m)}=\frac{U(\theta,\phi) }{r}\begin{cases}e^{i k(t+ r)}e^{i(m-p_i/2)\phi}\left(\begin{array}{c}
0\\
d^{j_i}_{-m,j_i}(\theta)
\end{array}\right)&p_i>0\\
e^{i k(t-r)}e^{i(m-p_i/2)\phi}\left(\begin{array}{c}
d^{j_i}_{-m,-j_i}(\theta)\\
0
\end{array}\right)&p_i<0
\end{cases}
}
which we can express terms of a 2D field $\chi_{\mu_i,m}(t,r)$:
\eq{
\psi_{\mu_i}(r,\theta,\phi)=\frac{U(\theta,\phi)}{r} \sum_m \left(e^{i (m-p_i/2)\phi}d^{j_i}_{-m,\pm j_i}(\theta)\right)\chi_{\mu_i,m}(t,r)~,
}
that lives on the half-plane with coordinates $(t,r)$ with a boundary at $r=0$.

Note that the fermions are polarized. This may seem like an artifact of restricting to fermions of finite energy density near the origin, however, we can see that this must be true from a simple scaling argument. In the low energy theory, the only mass is the cutoff scale of the IR theory (set by the $W$-boson mass). Since the 4D fermions $\psi_{\mu_i}$ have scaling dimension $\thalf$ and 2D fermions $\chi_{\mu_i}$ have scaling dimension $\half$, we see that the zero-modes must have scaling $\psi_{\mu_i}\sim \frac{1}{r}f(m_Wr)$ so that the higher angular momentum zero-modes are suppressed by the UV scale. It then follows that the tower of scattering states built on these zero-modes also do not contribute to the low energy effective theory.

Notice that this differs from the case of massive fermions which are studied in \cite{Moore:2014jfa}. In the massive case, the wave function is exponentially suppressed by $\frac{e^{-ur}}{r}$ for mass $u$. Because of this, the spectrum of low energy modes contains an entire tower of $L^2$-normalizable higher angular momentum states corresponding to the $c_{+,m}$-solutions in \eqref{abeliansolutions}. We believe that this modified spectrum of low energy modes will lead to interesting effects which we plan to address in a future paper.

\subsection{Effective 2D Theory and Boundary Conditions}

%non-confining in the IR and we can treat 
% 
% And further, there is no reason why the dynamical gauge field should be spherically symmetric. 
%
%However, the inclusion of  
%Because of this, it is simpler to understand the monopole-fermion interaction in terms of boundary conditions rather than in terms of scattering processes.
%  
% 
% And further, there is no reason why the dynamical gauge field should be spherically symmetric. 
% To discuss boundary conditions, we will take the viewpoint of  

The dynamics of the low energy dynamics of the fermion can now be described by an effective 2D theory of the $\chi_{\mu_i,m}$ fields.  Explicitly, if expand the fermion field operator 
\eq{
	   \psi_{\mu_i}(r,\theta,\phi)=\frac{U(\theta,\phi)}{r}\sum_m\int dk \,c_{i,m}^{(k)}\hat\psi_{\mu_i}^{(k,m)}(r,\theta,\phi)~,
} 
where the $c_{k,m}^{(i)}$ are creation operators, then we find that spherically reducing  $\psi_{\mu_i}$  results in the 2D field
\eq{
\chi_{\mu_i,m}(t,r)=\int dk\,\chi_{\mu_i,m}^{(k)}(t,r)=\int dk\, c_{i,m}^{(k)}\begin{cases}e^{ik(t+r)}\chi_+&p_i>0\\
e^{i k(t- r)}\chi_-& p_i<0
\end{cases}
}
which now lives 
on a half-plane with coordinates $(t,r)$ with a boundary at $r=0$ and where $\sigma^3\chi_\pm=\pm \chi_\pm$.  
Here we see that the fermion polarization in 4D is correlated with the chirality of the modes in the effective 2D theory. 

In other words, for each $\psi_{\mu_i}$ with $p_i>0$ we get $p_i$ chiral/left-moving 2D fermions corresponding to the $p_i$ components of the 4D spin-$\frac{p_i-1}{2}$ multiplet. Similarly, for each $\psi_{\mu_i}$ with $p_i<0$ we get $-p_i$ anti-chiral/right-moving 2D fermions corresponding to the $-p_i$ components of the 4D spin-$\frac{-p_i-1}{2}$ multiplet. %Additionally for fermions with $p_i=0$ the spherical reduction is equivalent to the restriction to the s-wave in the partial wave expansion which is unpolarized. 

We now see that the dynamics of the spherical fermion waves are described by left- and right-moving fermions (chiral/anti-chiral fermions) on the half plane. Since the fermions can reach the boundary, we must impose boundary conditions on the fields there that relate the left- and right-handed fermions.

The question of how to describe fermionic boundary conditions is somewhat subtle. One approach is to simply impose boundary conditions of the form 
\eq{
\psi_i=M_{ij}\tilde\psi_j\big{|}_{r=0}\quad \text{or}\quad \psi_i=M_{ij}\tilde\psi_j^\dagger\big{|}_{r=0}~,
}
where $\psi_i$ are chiral and $\tilde\psi_i$ are anti-chiral and $M_{ij}$ is an invertible matrix. However, a more general set of boundary conditions are given by the bosonizing the fields 
\eq{
\psi^\dagger_i\psi_i=J_i=\partial H_i\quad, \quad \tilde\psi_i^\dagger\tilde\psi_i=\bJ_i=\bar\partial \tildeH_i~, 
}
and then imposing boundary conditions on the bosonized fields $H_i,\tildeH_i$. Given a  mode expansion 
\eq{
H_i(z)=\sum_n H_{i,n}z^{n} \quad, \quad \tildeH_i(\bz)=\sum_n \tildeH_{i,n} \bz^n~,
}
the general boundary condition is given 
\eq{
H_{i,n}=\CR_{ij}\tildeH_{j,n}\big{|}_{r=0}~,
}
which can alternatively be written in terms of the currents
\eq{\label{curentbosonbc}
J_i=\CR_{ij}\bJ_j\big{|}_{r=0}~. 
}
For example, the standard Neumann and Dirichlet boundary conditions are given 
\eq{
{\rm Neumann}&:~ (\partial-\bar\partial)(H_i+\tildeH_i)\big{|}_{r=0}= 0\quad\,\Leftrightarrow\quad H_{i,n}=\tildeH_{i,n}\\
{\rm Dirichlet}&:~(H_i+\tildeH_i)\big{|}_{r=0}=0~\qquad\qquad\Leftrightarrow\quad H_{i,n}=-\tildeH_{i,n}
}
Then, by identifying the Neumann boundary condition with $(H_i-\tildeH_i)\big{|}_{r=0}$, we can apply the bosonization map to write the above boundary conditions in terms of the fermion fields:
\eq{
{\rm Neumann}:&~  H_i=\tildeH_i\big{|}_{r=0}\qquad\,\Rightarrow \quad \psi_i=\tilde\psi_i\big{|}_{r=0}~,
\\
{\rm Dirichlet}:&~H_i=-\tildeH_i\big{|}_{r=0}~\quad\Rightarrow \quad \psi_i=\tilde\psi_i^\dagger\big{|}_{r=0}~.
}
However, more general boundary conditions of the type \eqref{curentbosonbc} do not have such a simple interpretation in terms of UV fermion fields.

This formalism can be used to construct boundary conditions that preserve gauge and global symmetries \cite{Smith:2019jnh,Smith:2020rru,Smith:2020nuf}. 
Consider the case of $N$ chiral fermions $\{\psi_{i}\}$ and $N$ anti-chiral fermions $\{\tilde\psi_{i}\}$ in 2D on the half plane %Here we would like to consider unitary boundary conditions that relate... 
%
%We can classify the allowed boundary conditions by the preserved symmetries of the fermions. Thus, we like to consider 
with $N$ $U(1)$  symmetries indexed by $a=1,...,N$ under which the fermions have charges:
\begin{center}\begin{tabular}{c|c}
&$U(1)_a$\\
\hline
$\psi_{i}$& $Q_{ai}$\\
$\tilde\psi_{i}$& $\bar{Q}_{ai}$
\end{tabular}\end{center}
where we assume that the $U(1)_a$ are  independent (i.e. $Q_{ai},\bar{Q}_{ai}$ are invertible matrices)\footnote{ 
The condition that there are no anomalies among the $U(1)_a$ is given by 
\eq{
\sum_i Q_{ai}Q_{bi}-\bar{Q}_{ai}\bar{Q}_{bi}=0\quad\forall a,b~.
}
Here we will make no assumption about the $U(1)_a$ being non-anomalous. }.

For these fermions we can construct the fermion species currents
\eq{
J_{i}=\psi^\dagger_{i}\psi_{i}\quad, \quad \bar{J}_i=\tilde\psi^\dagger_{i}\tilde\psi_{i}~,
}
with which we can 
%use the $Q_{ai},\bar{Q}_{ai}$ to 
identify the $U(1)_a$ currents 
\eq{
\CJ_{a}=Q_{ai}J_i\quad, \quad \bar\CJ_a=\bar{Q}_{ai}\bar{J}_i~.
}

We would now like to construct a boundary condition that preserves the $U(1)_a$ symmetries. Such a boundary condition has the property 
\eq{\label{RQdef}
(\CJ_a-\bar\CJ_a)\Big{|}_{r=0}=0~.}
Then, using the fact that the $Q_{ai},\bar{Q}_{ai}$ are invertible, we can relate this to 
\eq{\label{bndstatesymcons}
(J_i-\CR_{ij}\bar{J}_j)\Big{|}_{r=0}
%|bnd\rangle
=0\quad, \quad \CR_{ij}=(Q^{-1})_{ia}\bar{Q}_{ja}~.
}
In the case where the symmetries are non-anomalous, $\CR_{ij}$ is a rational, orthogonal matrix. 
Physically, this means for any incoming fermion of fixed species $i$ the boundary will reflect to an out-going   collection of currents corresponding to the non-zero matrix components $\CR_{ij}$.

The Callan-Rubakov effect corresponds to a boundary conditions specified by  particular choice of $\CR_{ij}$.  In fact, the idea of bosonizing the spherical modes to obtain effective boundary conditions on the monopole goes all the way back to the original papers of Callan \cite{Callan:1982ac,Callan:1983tm}. 

Previously, only the boundary conditions for $N_f$ fundamental fermions in the presence of a minimal $SU(N)$ monopole were known\footnote{The boundary conditions for the minimal monopole in the $SU(5)$ Georgi-Glashow model is also known, but it is equivalent to the case of the minimal monopole in $SU(N)$ with fundamental fermions.} \cite{Callan:1982ac,Callan:1982ah,Callan:1982au,Callan:1983ed,Callan:1983tm,Rubakov:1982fp,Polchinski:1984uw,Maldacena:1995pq,Affleck:1993np}. 
%fermion boundary conditions on a monopole were those that describe the low energy effective theory of an $SU(N)$ gauge theory with $N_f $ fundamental fermions (which have $|q_i|=1,0$) in the presence of a minimal monopole. 
The low energy theory of this configuration is described by $N_f$ in-going and $N_f$ out-going s-wave fermions. These are described by an effective 2D theory of $N_f$ left- and $N_f$ right-moving fermions on the half-plane. In the above formalism, the monopole boundary condition 
is specified by 
\eq{\label{CRogR}
\CR_{ij}=\begin{cases}
1-\frac{2}{N_f}&i=j\\
-\frac{2}{N_f}&i\neq j
\end{cases}
}
Here we see that $\CR$ is very non-diagonal so that a generic scattering process will have  non-trivial components along all out-going fermion currents. This matrix $\CR_{ij}$ corresponds to the boundary condition that preserves a collection of non-anomalous symmetries with charge matrices given by 
\eq{
Q_{ai}=\begin{cases}
\delta_{a,i}-\delta_{a,i+1}&a<N_f\\
1&a=N_f
\end{cases}\quad, \qquad 
\bar{Q}_{ai}=\begin{cases}
\delta_{a,i}-\delta_{a,i+1}&a<N_f\\
-1&a=N_f
\end{cases}
}
Here, the $U(1)_1,..., U(1)_{N_f-1}$ symmetries are a labeling symmetry where $U(1)_a$ counts the flux of  $\psi_a,\tilde\psi_a$ minus the flux of $\psi_{a+1},\tilde\psi_{a+1}$ and $U(1)_{N_f}$ is the gauge charge. One can explicitly show that the matrix $\CR_{ij}$ in \eqref{CRogR} is given by the formula $\CR_{ij}=(\bar{Q}^{-1}Q)_{ij}$ using the above charge matrices.

As we show later, the physics of these boundary conditions comes from the fact that the fermions couple to the dynamical dyon degree of freedom (trapped W-bosons) on the monopole world volume. The boundary conditions then encode the physics of a fermion wave which, in scattering off of the monopole, excites the dyon degree of freedom which then radiates away the accumulated gauge charge and energy into all out-going modes that couple to it \cite{Polchinski:1984uw}. %This is the physics of the Callan-Rubakov effect.

At first glance, it appears strange that the out-going flux for each $\bJ_i$ is generically fractional. However, this is simply a fact that the currents $J_i,\bJ_i$ are not conserved currents in the quantum theory: each fermion number symmetry is broken by an ABJ anomaly. The magnetic field of the monopole together with the electric field of any low energy charged fermion combine together to turn on a non-trivial gauge background that breaks the conservation of all $J_i,\bJ_i$. The fractional flux is simply a reflection of the fact that species number is not a good quantum number in our theory. 
Rather, the boundary condition is diagonalized by taking a basis of conserved currents rather than a basis of anomalous species currents whose symmetries are broken in the presence of the monopole. It should not be a surprise that non-perturbative effects are reflected in monopole-fermion scattering even at low energy since monopoles are inherently non-perturbative features of non-abelian gauge theory.

The general effective boundary conditions for higher charge monopoles and/or for fermions in arbitrary representations are not known. It is presumed that the physics is very similar to the case of the standard Callan-Rubakov effect. However, this is not immediately obvious since increasing the magnetic field or dimension of the coupled representations naturally leads to fermions of higher angular momentum which experience an angular momentum barrier that repulses their wave function from the core of the monopole. 
 %, but as we will see in the preceding sections there are interesting new wrinkles to the story. 
Further, it is generally unknown how to characterize what kinds of symmetries are preserved by monopole boundary conditions although it is presumed that any symmetry which experiences an ABJ-type anomaly will be violated.  In the rest of this paper, we will derive the general low energy effective boundary conditions for spherically symmetric monopoles coupled to fermions of arbitrary representation and discuss their interplay with global symmetries.

\section{Monopole Lines from Smooth  Monopoles}

\label{sec:3}

Now we will discuss the general spherically symmetric monopole and derive the origin of their dyon degrees of freedom.  

Again, our setting is $SU(N)$ gauge theory with an adjoint-valued Higgs field $\Phi$ and a Weyl fermion $\psi_R$ that transforms in representation $R$ of $SU(N)$. We will consider a vacuum in which the Higgs field obtains a vev, $\Phi_\infty\in \fsu(N)$, which breaks
\eq{
SU(N)\to G_{\Phi_\infty}=\frac{U(1)^r\times \prod_a SU(N_a)}{\Gamma}\quad, \quad \Gamma\subseteq \prod_a \IZ_{N_a}~. 
}
Here we will denote the generators of $U(1)^r$ to be $\{h^I\}$. 

This theory has a spectrum of smooth monopoles which are finite energy, time independent gauge field configurations with vanishing electric field that solve the Bogomolny equation% \cg{Why BPS?}
\eq{
B_a=D_a \Phi~,
}
that have the asymptotic behavior
\eq{
\lim_{r\to \infty}\vec{B}\sim \frac{\gamma_m}{2r^2}\hat{r}+O(1/r^3)\quad, \quad \lim_{r\to \infty}\Phi=\Phi_\infty-\frac{\gamma_m}{2r}+O(1/r^{2})~,
}
where $\gamma_m=\sum_I n_I h^I$ $(n_I\in \IZ_{\geq0})$ is the magnetic charge. 

In the IR limit, these monopoles are well described by a monopole defect as we discussed in the previous section. We will be interested in spherically symmetric monopoles because 1.) they give rise to spherically symmetric boundary conditions in the IR and 2.) there is an explicit construction of their field configurations.

A large class of smooth, spherically symmetric non-abelian monopoles can be constructed by embedding the minimal $SU(2)$ monopole into higher rank non-abelian gauge theories.  However, this only a special class of spherically symmetric monopoles. %One class of monopole that does not fall into this classification are the general spherically symmetric monopoles. 
More generally, a monopole is spherically symmetric if it is rotationally invariant up to a gauge transformation. This is equivalent to the statement that the gauge field configuration is invariant under an augmentation of the 
standard angular momentum generators $\vL=-i \vec{r}\times \vec\nabla$ by generators of an $SU(2)$ subgroup $\vT$ (which satisfy $[T_i,T_j]=i\epsilon_{ijk}T_k$): % of the gauge group and then demand that the gauge field is invariant under the new generators 
$\vK=\vL+\vT$. Without loss of generality we we will restrict to the case where $\vT$ is the maximal, irreducible embedding. 

Here we will consider spherically symmetric $SU(N)$ monopoles and we will take the generators $\vT$ to define a subgroup $SU(2)_T\subset SU(N)$. The monopole connection can be written in a spherically symmetric gauge as 
\eq{
\vA=\frac{\hat{r}}{r}\times (\vec{\fM}-\vT)~, 
}  
where  $\vec{\fM}$ is a $\fsu(N)$-valued vector that satisfies
\eq{
[K_i,\fM_j]=i \epsilon_{ijk}\fM_k~. 
}
Such a $\fM_k$ can be constructed by taking the rotation of a $\theta,\phi$-invariant vector field along the $\hat{z}$-axis:
\eq{
\fM_i=\Omega\, M_i(r)\,\Omega^{-1}\quad, \quad \Omega=e^{-i \phi T_3}e^{- i \theta T_2}~, 
}
where $M_3(r)=0$ and $[T_3,M_\pm(r)]=\pm M_\pm(r)$ for $M_\pm=M_1\pm i M_2$. We will refer to this as the \emph{vector gauge}. 

Following \cite{Wilkinson:1978zh}, we can then use this ansatz to construct spherically symmetric $SU(N)$ monopoles. If we parametrize 
\eq{
&T_3=\half\diag(N-1,N-3,...,-N+1)~,\\
&
M_+(r)=\big(M_-(r)\big)^T=\left(\begin{array}{ccccc}
0& a_1(r)&&&\\
&0& a_2(r)&&\\
&&\ddots&\ddots&\\
&&&0& a_{N-1}(r)\\
&&&&0
\end{array}\right)~,\\&
\Phi\big{|}_{\hat{z}}=\Phi=\diag(\phi_1(r)\,,\,\phi_2(r)-\phi_1(r)\,,\,...\,,\,-\phi_{N-1}(r))~,
} 
where $\Phi(\vx)=\Omega \Phi\big{|}_{\hat{z}}\Omega^{-1}$, then the Bogomolny equation becomes 
\eq{
\frac{d\phi_I(r)}{dr}&=\frac{1}{r^2}\Big( a_I^2(r)-I(N+1-I)\Big)\\
\frac{d a_I(r)}{dr}&=\half\Big(2\phi_I(r)-\phi_{I-1}(r)-\phi_{I+1}(r)\Big) a_I(r)~,
}
which has smooth, finite energy solutions as constructed in \cite{Wilkinson:1978zh}. The solutions are specified by  an additional embedding $SU(2)_{I}\hookrightarrow SU(N)$ which is generated by $I_i\subset \fs\fu(N)$ ($[I_i,I_j]=i\epsilon_{ijk}I_k$) such that 
\eq{
&\lim_{r\to 0}M_\pm(r) =T_\pm+O(1)\quad, \quad \lim_{r\to \infty}M_\pm(r) =I_\pm+O(1/r^\delta)\quad\delta>0~,
}
and $\lim_{r\to0}\Phi(r)=0$. Such a choice of $\vI$ defines a monopole with magnetic charge\footnote{Here we use the convention where $e^{4\pi i I_k},e^{4\pi i T_k}=\mathds{1}_{SU(2)}$ while $e^{2\pi i \gamma_m}=\mathds{1}_{SU(N)}$. }
\eq{
\gamma_m=2(T_3-I_3)~,
}
 and Higgs vev\footnote{In the case where $\vT$ is not a maximal embedding in $SU(N)$, the Higgs vev can also have a constant term:
 \eq{
 \Phi_\infty=v\left(T_3-I_3\right)+\phi_\infty~,
 }
such that $[\vT,\phi_\infty]=[\vI^{(i)},\phi_\infty]=0$, $\forall i$. In this case, the Higgs vev is non-trivial in the monopole core:
\eq{
\lim_{r\to 0}\Phi(r)=\phi_\infty~. 
}
In this paper we will restrict to the case where $\phi_\infty=0$. \label{footnotevev}}
\eq{
\Phi_\infty=v\left(T_3-I_3\right)~,%+\phi_\infty~,
}
up to a choice of $v\in \IR$. % and $\phi_\infty\in \fsu(N)$.% such that $[\vT,\phi_\infty]=[\vI^{(i)},\phi_\infty]=0$, $\forall i$. In most of our discussion here we will consider $\phi_\infty=0$.  
Here we additionally assume that $[\vI,\gamma_m]=0$ and that  
the decomposition of $\vI$ into irreducible components $\vI=\bigoplus_{i}\vI^{(i)}$  
is non-trivial ($\vT\not\cong \vI$). % and that $I_\pm^{(i)}$ is expressible in terms of a linear combination of raising and lowering operators $E^\pm_I$ for simple roots.\footnote{In other words we restrict to $\vI_\pm$  such that $\lim_{r\to \infty}M_\pm(r)=I_\pm$ with the form of $M_\pm(r)$ specified above.} 

The above restrictions on $\vI$ imply that the magnetic charge and Higgs vev are constant along each irreducible component $\vI^{(i)}$ and hence the choice of $\vI$ defines the long range-non-abelian gauge field. If we denote the dimension of the irreducible component of $\vI^{(i)}$ as $N_i$, then the non-abelian part of the long range gauge symmetry of the theory is given by 
\eq{
G_{non\text{-}ab.}=\prod_{i\,|\,N_i\neq1}SU(N_i)~.
}

The gauge field for the spherically symmetric monopole is also expressible in a more convenient gauge:
\eq{\label{Acan}
A_{can.}=2 T_3A_{Dirac}+i\frac{ M_+(r)}{2}e^{-i \phi}(d\theta-i \sin\theta d\phi)-i \frac{M_-(r)}{2} e^{i \phi}(d\theta+i \sin\theta d\phi)~,
}
We will refer to this as the \emph{canonical gauge} which is related to the original gauge by $g=e^{-i {\phi} T_3}e^{- i {\theta} T_2}e^{i {\phi} T_3}$. In this gauge, we can explicitly can compute the field strength:
\eq{
F 
=&\left(T_3-\frac{[M_+(r),M_-(r)] }{2}\right)d^2\Omega\pm  \ihalf M'_\pm(r) e^{\-m i \phi}dr\wedge (d\theta\mp i \sin\theta d\phi)~, 
}
from which we can read off the magnetic field along the $\hat{z}$-axis:
\eq{
B_r=\frac{2T_3-[M_+(r),M_-(r)]}{2r^2}\quad, \quad B_\pm=\frac{1}{r}\partial_r M_\pm(r)~.
} 
%We now see that normalizability of the gauge field and the asymptotic magnetic field fixes the asymptotic behavior of $M_\pm(r)$:

%Additionally, the Higgs field has the limiting behavior \cite{Wilkinson:1978zh}
%\eq{
%\lim_{r\to 0}\Phi(r)=\phi_\infty~,
%}
%which again, we will take to be zero for most of our discussion -- the modifications to the story for $\phi_\infty\neq 0$ are straightforward. 

The $ a_I$ parameterizing $M_\pm(r)$ are divided into two different classes by their asymptotic behavior. If $\lim_{r\to \infty} a_I(r)=0$, then it has the asymptotic behavior
\eq{
\lim_{r\to \infty} a_I(r)\sim e^{-\Delta_I m_Wr}\quad, \quad \Delta_I=\frac{2m_I-m_{I-1}-m_{I+1}}{2}~,
}
%then 
%\eq{
%[E^+_I,I_-]\neq 0\quad\Rightarrow \quad \lim_{r\to \infty} a_I>0~, }
%whereas
%\eq{\label{Iexploc}
%[E^+_I,I_-]=0\quad \Rightarrow \quad \lim_{r\to \infty} a_I\sim e^{-\Delta_I\,  r}\quad, \quad \Delta_I=\frac{2m_I-m_{I-1}-m_{I+1}}{2}~,
%}
where $\gamma_m=\sum_I m_I H_I$ and $\{H_I\}$ are a basis of simple coroots such that $[H_I,E^\pm_J]=\pm C_{IJ}E^\pm_J$ and $m_I\in \IZ_{\geq0}$.\footnote{
Here we
 pick  a basis  $\{H_I\}$ of the Cartan of $SU(N)$ which in the fundamental representation is given by:
\eq{
R_f(H_I)_{ij}=\delta_{iI}\delta_{jI}-\delta_{i,I+1}\delta_{j,I+1}~.
}
In this basis we can express  % \cg{was other basis better?}
\eq{
T_3=\half\sum_{I=1}^{N-1}I(N-I)H_I~, %\quad \Phi=\sum_I H_I \Phi_I\quad, \quad \Phi_I=\phi_I-\phi_{I-1}~.
}
for the maximal embedding in $SU(N)$. 
} Note that our construction ensures $\Delta_I\geq0$. 
Physically, this means that along broken directions in the gauge group, the W-bosons are  exponentially confined to the monopole core as in the case of the minimal $SU(2)$ monopole. 

In our following analysis we will also need the $m_W$ dependence of the solution. % \footnote{Here the $W$-boson mass is defined independently for each diagonal block of $SU(N)$ which corresponds to the irreducible components of $\vT$. Assuming $\phi_\infty=0$, the $W$-boson mass for the block defined by $\vT^{(i)}$ is  $m_W=v_i$.}
 In particular, since $A_{can.}=A_\mu dx^\mu$ in equation \eqref{Acan} is dimensionless,  $a_I(m_Wr)$ is a dimensionless function of the dimensionless combination $m_Wr$. 
%\eq{
%\alpha_I(r)=m_W\,a_I(m_Wr)~,
%}
%where $a_I(m_Wr)$ is a dimensionless function of $m_Wr$. 

\subsection{IR Limit}

Now let us study these monopoles in the IR limit.  
To do so, we 
 restrict to energies  $E/m_W<<1$ which can be achieved by effectively sending $m_W\to \infty$ while keeping $E$ fixed. This scaling affects the form of the gauge field by projecting to the leading term in an asymptotic expansion in powers of $1/m_Wr$:
\eq{
\lim_{m_W\to \infty}A_{can.}&=\lim_{m_W\to \infty}2 T_3A_{Dirac}\pm i\frac{M_\pm (m_Wr)}{2}e^{\mp i \phi}(d\theta\mp i \sin\theta d\phi)\\
&=2 T_3A_{Dirac}\pm i\frac{I_\pm }{2}e^{\mp i \phi}(d\theta\mp i \sin\theta d\phi)+O(1/m_Wr)\\
&%\underset{g_I}{
\sim
%}
 \gamma_m A_{Dirac}+O(1/m_Wr)~,
}
where the final step is equivalence up to a gauge transformation by 
$g_I=e^{-i {\phi} I_3}e^{- i {\theta}I_2}e^{i {\phi} I_3}$. 
Thus we see in the IR that 
%limit $m_W\to \infty$ that 
the non-abelian monopole configuration  approaches the abelian monopole.

Now we would like to describe the low energy effective gauge theory in the presence of 
%action for spherically reduced fermions in the presence of 
a singular monopole that arises from a $SU(N)$ gauge theory.    
As is standard in spontaneous gauge symmetry breaking, the low energy gauge bosons can be described by propagating vector fields for the gauge group $G_{\Phi_\infty}$. The gauge fields along the directions in $\fsu(N)$ that do not commute with $\Phi_\infty$ are massive away from the monopole. However, since $\lim_{r\to 0}\Phi=0$ we see that the asymptotically massive gauge fields become massless inside the core of the monopole. In essence, the non-trivial Higgs profile acts as a potential well that traps the broken gauge degrees of freedom in the monopole core. 

We can now parametrize the perturbations of the gauge field around this background: 
\eq{
A_{SU(N)}=A_{G_{\Phi_\infty}}+2 T_3A_{Dirac}+(W_++W_-)~,
}
in which we separate the degrees of freedom that obtain a mass from the Higgs field from those that are free-propagating non-abelian degrees of freedom. Here we distinguish between three components: 1.) $A_{G_{\Phi_\infty}}$ are the long range propagating gauge fields of $G_{\Phi_\infty}$, 2.) $2 T_3 A_{Dirac}$ is the abelian component of the classical solution. %We have separated the abelian component of the classical solution is along the $G_{\Phi_\infty}$ direction in $SU(N)$. 
and 3.) the $W_+,W_-$ encode the broken directions in $SU(N)$. The broken directions in $SU(N)$ can be further parametrized as 
\eq{\label{gaugefieldfields}
W_\pm=\ihalf e^{\pm i \sum^\prime_I\varphi_I h^I}M_\pm(r)e^{\mp i \phi}(d\theta\mp i \sin\theta d\phi )+w_\pm (x)~,
}
where $W_+=(W_-)^\dagger$ and the sum $\sum^\prime_I$ runs over the generators of $U(1)^r$ in $G_{\Phi_\infty}$.  Here $\varphi_I(x)$ are periodic-valued dynamical fields that parametrize the phase of the $ a_I(r)$ for which $\lim_{r\to \infty} a_I=0$ 
and $w_\pm$ are complex 1-form fields that take value in the broken directions in $SU(N)$. 

Plugging this parametrization of the gauge field into the action, we find that the $w_\pm$ degrees of freedom obtain a mass:
\eq{
\CL&=...+\half \Big|[W^+,\Phi]\Big|^2+\frac{1}{2g^2}\Big|[W^+,W^-]\Big|^2=...+\half \Big{|}[w_+(x),\Phi]\Big{|}^2+\ghalf \Big{|}[w_+(x),M_-]\Big{|}^2\\
&=...+\half m^2(x)|w_+|^2(x)~. 
}
Here we see that the spatially dependent mass has two components. The first comes from the non-trivial Higgs vev which gives a mass away from the core. The second comes from the profile of $M_\pm(r)$ which contributes to the mass inside the core\footnote{The contribution of $M_\pm(r)$ to the mass is always non-trivial due to the fact that $\vT$ generate the maximal $SU(2)_T$ embedding.}. Thus, all $w_\pm$ acquire a mass of order $m_W$ and  are completely frozen out in the IR.  `

Thus, the $\varphi_I$, which are massless, are the only dynamical degrees of freedom from the broken directions in $SU(N)$. 
%On the other hand, the $\varphi_I$ are not massive. 
However, since the  $\varphi_I$ are phases of the $ a_I(r)$ that exponentially decay away from the monopole core, the $\varphi_I$ are themselves confined to the monopole core.  This is clear from their kinetic term that comes from the kinetic term for the gauge field:
\eq{
\Tr\big\{(\partial_t W_+)(\partial_t W_-)\big\}=\sum_I a_I^2(r)  (\dot\varphi_I)^2~.
}
In the limit $m_W\to \infty$, we see that the above kinetic term is approximately
\eq{
\Tr\big\{(\partial_t W_+)(\partial_t W_-)\big\}\sim \sum_I  e^{-2 \Delta_I m_W r} (\dot\varphi_I)^2~.
}
Again, we see that this is peaked near $r=0$ and the integral scales like $1/m_W$ which leads to a kinetic term 
\eq{
S_{kin}=\int \frac{1}{2 }\dot\varphi_I^2~ \delta (m_W r)dr\,dt=\int \frac{\dot\varphi_I^2 }{2m_W}\delta(r)dr\,dt~.
}
Here $\varphi_I\sim \varphi_I+2\pi $ is a dimensionless field. The fact that the kinetic term has a prefactor $1/m_W$ implies that we should interpret $\varphi_I$ as the angular coordinate for a particle of mass $m_W$ that is moving on a circle of radius $1/m_W$. This can be seen from rescaling $\varphi_I\to m_W \varphi_I$ while changing the periodicity $\varphi_I\longmapsto \varphi_I+2\pi /m_W$:
\eq{
S_{kin}\longmapsto\int  \half m_W^2 \dot\varphi_I^2~ \delta(m_Wr)dr\,dt=\int\half m_W \dot\varphi_I^2(t)dt~.
} 
%
%Now we would like to discuss a subtle point. 
In addition to the kinetic term for $\varphi_I$, the kinetic term of the $SU(N)$ gauge field also contains a coupling to the bulk gauge fields:
\eq{
\Tr\big\{\partial_t W_+ [A_0,W_-]\big\}=  a_I^2(r) \dot \varphi_I A_0~,
}
where here $A_0$ is the bulk $U(1)$ gauge field under which $\varphi_I$ is charged. This leads to a coupling term in the quantum mechanics 
\eq{
S_{A_0\,}=\int A_0\, \dot\varphi_I ~\delta(m_W r)dr\,dt\longmapsto\int A_0\, \dot \varphi_I \delta(r)dr\,dt~,
}
where here we have rescaled $\varphi_I$ as above. 

This coupling induces an electric source term in the bulk equations of motion 
\eq{
\nabla \cdot E \approx\dot\varphi_I \delta^{(3)}(x)~,
}
which means that $\varphi_I$-excitations elevate the monopole to a dyon.  The contribution of the long range electric field to the energy is of order of the cutoff scale which is of order of $m_W$. This implies that although $\varphi_I$ is a light field in the sense that its kinetic term is suppressed by $1/m_W$, unlike the $w_\pm $ fields. However,  the fluctuations of $\varphi_I$ source long range electric field that carry a large amount of energy. This implies that the $\varphi_I$ constitutes a dynamical quantum degree of freedom, but its coupling to the bulk gauge fields energetically forbids long lived excitations.

\subsubsection{Example: $SU(2)$ Monopole}

Now let us explicitly demonstrate the existence of the dyon degree of freedom in the low energy effective theory of $SU(2)$ gauge theory in the minimal monopole background. 

Here we parametrize the Lie algebra $\fsu(2)$ by $\{H,E^\pm\}$ which satisfy 
\eq{
[H,E^\pm]=\pm 2E^\pm \quad, \quad [E^+,E^-]=H~.
}
The monopole field configuration is explicitly given by 
\eq{
A_{SU(2)}&=H A_{Dirac}\pm \ihalf \left(\frac{m_Wr}{\csch(m_Wr)}\right)e^{\mp i \phi}(d\theta- i \sin\theta d\phi)E^\pm~,\\
\Phi&=m_W (\coth(m_Wr)-1/r)H~.
}
Now we will parametrize the dynamical fields as 
\eq{
A_{SU(2)}&=H(A_{Dirac}+\hat{A}_{U(1)})+E^+\left(\hat{w}_++\ihalf \left(\frac{m_Wr}{\csch(m_Wr)}\right)e^{- i \phi+i \hat\varphi}(d\theta\mp i \sin\theta d\phi)\right)+c.c.~,\\
\Phi&=H\left(\hat\rho+m_W \left(\coth(m_Wr)-\frac{1}{m_Wr}\right)\right)+\hat\phi_+E^++c.c.~,
}
where all dynamical fields are hatted. 

Now let us plug in our field parametrization into the action 
\eq{
S=\int \frac{1}{4}\Tr\big\{F_{\mu\nu}F^{\mu\nu}+\half |D_\mu \Phi|^2+\lambda [\Phi^\dagger,\Phi]^2\big\}~d^4x~.
}
Now we see that $\hat\phi_\pm$ from the quartic $\Phi$-interaction and coupling to the $W$-bosons and  $\hatw_\pm$ get a mass from the kinetic terms for $A_{SU(2)}$ and $\Phi$:
\eq{
m_{\phi_\pm}^2,m_{w_\pm}^2=m_W^2 r^2 \left(\coth(m_Wr)-\frac{1}{m_Wr}\right)^2+m_W^2 r^2 \csch^2(m_Wr)\gtrsim O (m_W^2)~.
}
Thus, these fields are integrated out and the $\hatw_\pm$ boson eats the Goldstone mode of $\hat\phi_\pm$. We now find that the low energy theory is described by the radial mode $\hat\rho$ (which is not important for our story), the long range $U(1)$ gauge field $A_{U(1)}$, and the dyon mode $\hat\varphi$. 

The only terms that arise in the action that involve $\varphi$ arise from the derivatives in the kinetic term:
\eq{
S_{\varphi}&=\int \frac{1}{8\pi}\Tr\Big\{\partial_\mu W_+\partial^\mu W_-+2 g [A^\mu,W_+]\partial_\mu W_-\Big\}d^4x\\
&=\frac{1}{4\pi}\int m_W^2r^2 \csch^2(m_W r)\left(\half (\partial_\mu \varphi)^2+2 g A ^\mu\partial_\mu \varphi\right)d^2\Omega \,dr \,dt~,
}
where $A=A_{U(1)}$. 
In the IR limit $m_W\to \infty$, we find that $m_W^2 r^2\csch^2(m_W r)\to \delta(m_Wr)$ and all spatial derivatives of $\varphi$ are suppressed. This leads to a contribution to the low energy effective action 
\eq{
S_{\varphi}=&\frac{1}{4\pi}\int  m_W^2r^2 \csch^2(m_W r)\left(\half \dot\varphi^2+g A_0\dot\varphi\right)d^2\Omega\, d^2x\\
&= \int \delta(m_Wx)\left(\dot \varphi^2+ g A_0\dot\varphi\right)d^2x=\int \left(\frac{1}{2}m_W\dot\varphi^2+g A_0 \dot\varphi\right)dt~,
}
where in the final step we rescaled $\varphi\to m_W\varphi$. Here we see that the dyon degree of freedom contributes to the effective low energy theory as a quantum mechanical rotor localized on the monopole world volume. A similar computation holds for the case of dyon degrees of freedom on a general spherically symmetric monopole.

\section{Low Energy Fermions}

\label{sec:4}

Now we would like to derive the interaction between the low energy fermion modes and the dyon degrees of freedom. Due to the fact that the long range electric field suppresses long lived excitations of the dyon field, any fermion-dyon interaction should be integrated out to give an effective boundary condition for the long range fermion fields. 

One way to determine the  fermion-dyon interaction is from the explicit form of the fermion zero-modes of the full UV Dirac equation. In solving the Dirac equation, we are treating the gauge field as a classical background so that the solutions encode the tree level interaction of the fermions with the monopole. We can then determine the dyon interaction by matching the low energy fermions onto the   large $r$ limit of the fermion zero-modes and imposing conservation of gauge symmetry.
%Then, from the fact that the low energy fermions are simply radial scattering modes which we can match onto the  the large $r$ limit of the fermion zero-modes, we can read off the  dyon interaction term from the large $r$ limit of the fermion zero-modes and by imposing conservation of gauge symmetry.  

\subsection{Fermion and Angular Momentum}

We would now like to solve for the fermion zero-modes in the presence of a spherically symmetric monopole. The spherical symmetry implies that the fermions can also be decomposed into representations of the total angular momentum generators which in the vector gauge are written:
\eq{
\vJ=\vL+\vS+\vT=-i \vec{r}\times \vec\nabla+\half \vec\sigma+\vT~
}
which generate the group $SU(2)_J$. 

Let us consider a Weyl fermion $\psi_R$ that transforms under $SU(N)$ with representation $R$.  
Since $\vJ$ is the sum of three commuting generators, we find that the $SU(2)_J$-representations will be composed of the tensor products of $\vL,\vS,\vT$. In general, this leads to representations of $SU(2)_J$ with non-trivial multiplicity for fixed $R$ due to the fact that generically many $SU(2)_L$ representations can produce a $SU(2)_J$ representation of fixed spin. 

To solve the Dirac equation, we will both gauge transform from the vector to the canonical gauge 
and implement a frame rotation (See  \eqref{framerot}). These transformations cause the rotation generators to transform  due to the fact that $\vec\nabla,\vT,\vS$ are not gauge/frame  invariant: 
\eq{
\vJ=-i \vr\times \left(\vec\partial+2i T_3\vec{A}_{Dirac}-i\sigma^3\half \cot\theta\hat\phi\right)+\left(T_3+\half\sigma_3\right)\hat{r}~.
}

As discussed in \cite{Brennan:2021ucy}, we can construct the representations of $SU(2)_J$ by diagonalizing $J^2,J_3$ which have the explicit forms: %$J_3=i \partial_\phi+\half T_3+\half\sigma_3$ and 
\eq{
J^2&=-\half\left[D^-_{T_3+\frac{\sigma^3}{2}+1}D^+_{T_3+\frac{\sigma^3}{2}}+D^+_{T_3+\frac{\sigma^3}{2}-1}D^-_{T_3+\frac{\sigma^3}{2}}\right] 
+\left(\frac{1}{2}\sigma_3+T_3\right)^2~,\\
J_3&=-i( \partial_\phi+iT_3)~,
}
where 
\eq{
D^\pm_q=\partial_\theta\mp \frac{1}{\sin\theta}\Big(i(\partial_\phi+i T_3)+q \cos\theta\Big)~.
}

To simultaneously diagonalize these,  consider a decomposition of the representation $R$ with respect to irreducible representations of $SU(2)_T$:
\eq{
R=\bigoplus_A R_A\quad\Rightarrow \quad \psi_R=\sum_A\psi_{R_A}~. 
} 
The basis of the representation space  $V_{R_A}$ can be indexed by weights
\eq{
V_{R_A}={\rm span}_\IC\{v_{\mu_a}\}\quad, \quad \mu_a\in \Delta_{R_A}~,
}
where $\Delta_{R_A}$ is the weight space of the representation $R_A$. 
Now if we decompose 
\eq{
\psi_{R_A}=\sum_a \psi^a_{R_A} v_{\mu_a}~,
}
so that the $\psi^a_{R_A}$ have charges $q_a$ where 
\eq{
{R_A}(T_3)\cdot v_{\mu_a}=q_a v_{\mu_a}\quad,\quad \langle T_3,\mu_a\rangle=\frac{q_a}{2}~, 
}
then we find that the eigenfunctions of $J^2,J_3$ are given by 
\eq{\label{psiparam}
\psi^{a\,(j)}_{R_A}=e^{i(m-q_a/2)\phi}\left(\begin{array}{c}f_{a,+}(r)\, d^{j}_{-m,\frac{q_a+1}{2}}(\theta)\\
f_{a,-}(r)\, d^{j}_{-m,\frac{q_a-1}{2}}(\theta)
\end{array}\right)~,
}
with quantum numbers 
\eq{
J^2\psi^{(j)}_{R_A}=j(j+1)\psi^{(j)}_{R_A}\quad, \quad J_3\psi^{(j)}_{R_A}=m\psi^{(j)}_{R_A}~,
}
where $d^{j}_{m,q}(\theta)$  are small Wigner $D$-functions. 
These representations are restricted by the fact that $d^{j}_{m,q}(\theta)$ is only well defined for $|m|,|q|\leq j$ and $m-j,q-j\in \IZ$. These restrictions imply that the allowed set of $j,m$ are given by 
\eq{
j&=\frac{2q_{{R_A}}+1}{2}+n_j\geq0\quad, \quad n_j \in \IZ~,\\
m&=-j,-j+1,...,j-1,j~.
}
where the representation $R_A$ is a spin-$q_{{R_A}}$ representation of $SU(2)_T$. Note that when $n_j<0$ that we must impose additional restrictions on the $f_{a,\pm}(r)$:%,f_{a,-}(r)$:
\eq{
f_{a,\pm }(r):=0~\text{ if } ~\left|\frac{q_a\pm 1}{2}\right|>j~.% \quad 
%f_{a,-}(r):=0~\text{ if }~\left|\frac{q_a-1}{2}\right|-j<0\\
}
which give  $SU(2)_J$ representations  a nested structure. To facilitate this we will define the spin-$j$ restricted weight space
\eq{
\Delta_{R_A}^{(j)}=\left\{\mu_a\in \Delta_{R_A}~\Big{|}~|\langle \mu_a,T_3\rangle|\leq j+1/2\right\}~,
}
which is simply the representation $R_A$ restricted to the subspace with non-trivial $\psi_{R_A}^{(j)}$. In other words, the spin-$j$ mode of $\psi_{R_A}$ is restricted to the component fields $\psi_{R_A}^a$ such that $\left|\frac{q_a}{2}\right|<j$.  
 
One way to make use of the representations of angular momentum is in solving for the fermion modes in abelian monopole background. In order to make contact with the representation theory above, we will gauge transform the standard abelian monopole background:
\eq{
A_{ab.}=\gamma_m A_{Dirac}~,%+O(1/r)~,
}
by $g_I=e^{-i {\phi} I_3}e^{ i {\theta}I_2}e^{i {\phi} I_3}$. 
to the equivalent of the canonical gauge:
\eq{
A_{can.}=2 T_3 A_{Dirac}\pm i \frac{I_\pm}{2}e^{\mp i \phi}(d\theta \mp i \sin\theta d\phi)~.
}
This is similar to the full non-abelian gauge field in the previous section except that $M_\pm(r)$ is replaced by its asymptotic value $M_\pm(r)\to I_\pm$ which encodes the long range abelian monopole. Note that this canonical abelian gauge is a good gauge in the IR theory with gauge group $G_{\Phi_\infty}$. 

The zero-modes of the abelian Dirac operator in the canonical gauge are solved for in \cite{Brennan:2021ucy}. They are somewhat cumbersome due to the fact that they involve Clebsh-Gordon coefficients, but can be simplified by a coordinate independent gauge transformation. Adapting the computation, we find that the spectrum of the Dirac operator can be written 
\eq{
\psi_{\mu_a}^{(k,m)}=\frac{U(\theta,\phi)}{r}e^{i(m-q_a/2) \phi}\begin{cases}
e^{ik(t+r)}\left(\begin{array}{c}
0\\ d^{j_a}_{-m, j_a}(\theta)
\end{array}\right)&q_a>0\\
e^{ik(t-r)}\left(\begin{array}{c} 
d^{j_a}_{-m,-j_a}(\theta)\\0
\end{array}\right)&q_a<0
\end{cases}\qquad j_a=\frac{|q_a|-1}{2}
}
where again $ d^j_{m,q}(\theta)$ is a small Wigner-$D$ function. 

In this gauge, we find that the low energy fermion modes can again be described by the 2D fields 
$
\chi_{\mu_a,m}(t,r)$ 
corresponding to the expansion 
\eq{
\psi_{\mu_a}=\frac{U(\theta,\phi)}{r}\sum_{m=-j_a}^{j_a} \left(e^{i (m-q_a)\phi}d^{j_a}_{-m,j_a}(\theta)\right)\chi_{\mu_a,m}(t,r)~.
}
See Appendix \ref{app:A} for more details. 

 \subsection{Fermion Zero-Modes}

Using the parametrization of a generic $SU(2)_J$ representation in \eqref{psiparam} acting on a $SU(2)_T$ representation $R_A$,  the smooth spin-$j$ solutions of the Dirac equation are given by  
\eq{\label{nonabeliansol}
\psi^{a}_{R_A}=e^{i(m-q_a/2)\phi}\left(\begin{array}{c}f_{a,+}(r)\,d^{j}_{-m,\frac{q_a+1}{2}}(\theta)\\
f_{a,-}(r)\,d^{j}_{-m,\frac{q_a-1}{2}}(\theta)
\end{array}\right)~,
}
where the functions $f_{a,\pm}(r)$ are given by 
\eq{
F(r)=P\,\exp\left\{-\int \CD(r)dr\right\}\,F_0\quad, \quad F(r)=\sum_{\mu_a\in \Delta_{R_A}^{(j)}}\left(\begin{array}{c}
f_{a,+}(r)\\
f_{a,-}(r)
\end{array}\right)\, v_{\mu_a}~,
}
for 
\eq{
\CD(r)=-\sqrt{(j+1/2)^2-\frac{q_a^2}{4}}\sigma^1 +\sigma^-M_++\sigma^+M_-+1~,
}
and $F_0$ a constant vector. As proven in \cite{Brennan:2021ucy}, there exists a unique vector $F_0$ that corresponds to a smooth, plane-wave normalizable solution of spin-$j$: 
\eq{
\lim_{r\to \infty}r \psi_{R_A}<\infty~,
}
of the time-independent Dirac equation iff the top and bottom weights of the spin-$j$ representation of $R_{A}$: 
  \eq{
\langle \mu_{top},T_3\rangle=-\langle \mu_{bot},T_3\rangle=2j+1\quad, \quad \mu_{top},\mu_{bot}\in \Delta^{(j)}_{R_A}~,
}
have charge of opposite sign under $\gamma_m$:
\eq{
\langle \mu_{top},\gamma_m\rangle\langle \mu_{bot},\gamma_m\rangle<0~.
} 
In total, one finds that  a fermion $\psi_R$ in representation $R$ of $SU(N)$ has a total number of fermion zero-modes:
\eq{
\dim_\IC\Big[\ker[\bar\sigma^\mu D_\mu]\Big]=\lim_{\epsilon\to 0^+}\sum_{\mu \in \Delta_R}n_R(\mu)\langle \mu, T_3\rangle~sign(\langle \mu-\epsilon\,w(\mu),T_3-I_3\rangle)~,
}
which decompose as $SU(2)_J$ representations as
\eq{
\psi_R\to \bigoplus_{\mu \in S_R}[{j_\mu }]^{\oplus n_R(\mu)} \quad, \quad j_\mu=\frac{\langle \mu,T_3\rangle-1}{2}~,}
where $[j]$ is the spin-$j$ representation and the sum runs over the set of weights:
\eq{
S_R=\Big\{\mu \in \Delta_R^+~\Big{|}~\langle \mu, T_3-I_3\rangle\times \langle w(\mu),T_3-I_3\rangle)<0\Big\}~,
}
where $w$ is the unique element of the Weyl group of $\fsu(2)_T\subset \fsu(N)$ 
 Here we will also use the notation $\mu^\ast=w(\mu)$. Note that $\mu_{top}^\ast=\mu_{bot}$. 

In order to match onto the IR fermion modes, we will also need the explicit asymptotic behavior of the zero-mode solutions. 
%in the limit $r\to \infty$.
 Let us consider the spin-$j$ zero-mode of $\psi_{R_A}$.  The spin-$j$ representation has a highest and lowest weight $\mu_{top},\mu_{bot}\in \Delta^{(j)}_{R_A}$. % which are acted on by  embeddings $\vI^{(top)}, \vI^{(bot)}$. %We will denote the restriction of $R_A$ to a particular $\vI^{(a)}$ embedding $R_{A,I^{(a)}}$. 
Up to coordinate independent gauge transformations, the spin-$j$ zero-mode of $\psi_{R_A}$ has asymptotic behavior
%In this limit the spin-$j$ zero-mode of $\psi_{R_A}$ is given by 
\eq{
\lim_{r\to \infty}\psi_{R_A}\sim\frac{1}{r}\left(\psi_{R_A}^{+}v_{\mu_{top}}-\psi_{R_A}^{-}v_{\mu_{bot}}\right)+O(1/r^{1+\delta})\quad, \quad \delta>0~,
}
where 
\eq{
\sigma^0\bar\sigma^r\psi_{R_A}^{\pm}=\pm\psi_{R_A}^{\pm}~,%\quad,\quad \sigma^0\bar\sigma^r\psi_{R_A}^{bot}=-\psi_{R_A}^{bot}~,
}
are spin polarized with respect to the radial direction. 
These $\psi_{R_A}^{\pm}$ are the top/bottom components of the spin-$j$ representation and hence match onto the zero-momentum scattering modes of $\chi_{\mu_{top/bot},m}(t,r)$ in the IR theory. See Appendix \ref{app:A} for more details. 
 
Interestingly, in the core of the monopole, the solutions to the Dirac equation have the form 
\eq{
\lim_{r\to 0}\psi_{R_A}= \sum_\ell r^{\ell}\psi_{R_A}^{(\ell)}\quad, \quad J^2 \psi_{R_A}=j(j+1)\psi_{R_A}~,
}
where $\psi_{R_A}^{(\ell)}$ is the spin-$j$ representation with orbital angular momentum $\ell$. 
The minimum $\ell$ given a fixed $j,R_A$ is 
\eq{
\ell_{min}=dim[R_A]-j-\half~,
}
which implies that fermion zero-modes generically vanish at $r\to 0$ except for  solutions with spin-$j=dim[R_A]-\half$.  

This leads to a confusing point. It is often said that the only the spin-$j=0$ modes of the fermion interact with the dyon degrees of freedom because their wave function has non-trivial support in the core. However, we see here that this is not exactly true: the modes with spin-$j=dim[R_A]-\half$ have no angular momentum barrier -- this only corresponds to the s-wave in the case of a minimal monopole and fermions of sufficiently small representation. 

In the next section we will show that the emphasis on the angular momentum repulsion from the core is a red herring: even plane-wave normalizable the modes that vanish at the origin still interact the the dyon degrees of freedom. %This is perhaps not too surprising given that the long range components of the fermion zero-modes do not have the same gauge charge. This implies that  the corresponding IR modes must be related at the core by a charged field on/near the monopole which is naturally fulfilled by the dyon degree of freedom. 
 
\subsection{Low Energy Effective Action}

Now we would like to use the explicit form of the fermion zero modes to derive the fermion-dyon coupling. As we discussed in Section \ref{sec:2}, the low energy limit of the fermions are described by the effective 2D fields $\chi_{\mu,m}(t,r)$. 

From the asymptotics of the fermion zero-modes in the previous section, it is clear that the long range degrees of freedom  are related by a boundary condition
\eq{\chi_{\mu,m}(t,r=0)\leftrightarrow \chi_{\mu^\ast,m}(t,r=0)~,
}
where $\chi_{\mu^\ast,m}=\chi_{w(\mu),m}$ is the charge conjugate of $\chi_{\mu,m}$ with respect to $SU(2)_T$. 

However, the boundary condition cannot be simply $\chi_{\mu,m}(t,r=0)= \chi_{\mu^\ast,m}(t,r=0)$ 
due to the fact that it does not preserve gauge invariance under the IR gauge group. To fix this, the boundary condition must be supplemented by a charged degree of freedom in the core. Since the 4D fermions only couple to the gauge field, this charged degree of freedom must come from the broken $W$-bosons. And further, as we demonstrated in the previous section, the only light charged degree of freedom in the core is the dyon degree of freedom. This implies that the boundary condition must be of the form 
\eq{
\chi_{\mu,m}(t,r=0)=e^{i c^I_\mu\varphi_I  }\chi_{\mu^\ast,m}(t,r=0)~,
}
where $c^I_\mu$ is a collection of integers such that the above boundary conditions are gauge invariant. 
This boundary condition can be encoded by the boundary term:
\eq{\label{dyonint1}
L_{bnd}=\sum_\mu\int \delta(r) e^{i c^I_\mu\varphi_I  }\chi_{\mu,m}^\dagger\chi_{\mu^\ast,m} \,d^2x+c.c.~. 
}

Let us now show how this boundary condition arises. As it turns out, this boundary condition  comes from a tree level interaction. If we expand the asymptotic form of the spin-$j$ 4D fermions in $SU(2)_J$-representations
\eq{
\psi_{R_A}^{(j),a}=\frac{U(\theta,\phi)}{r}\sum_m e^{i(m-q_a/2)\phi}\left(d^j_{-m,\frac{q_a+1}{2}}(\theta)\chi^{(j),+}_{\mu_a,m}(t,r)+d^j_{-m,\frac{q_a+1}{2}}(\theta)\chi^{(j),-}_{\mu_a,m}(t,r)\right)~,
}
where $\chi^+_{\mu_{top},m},\chi^-_{\mu_{bot},m}=0$, then we find upon plugging into the action with the low energy $W$-boson degrees of freedom \eqref{gaugefieldfields} that the couplings to $\varphi_I$ are of the form 
\eq{
L_{bnd}=\sum_{I,\mu_i,m}\int \alpha_{I}(r)e^{i\varphi_I}(\chi^{(j),+}_{m,\mu_i})^\dagger \chi^{(j),-}_{m,\mu_i-\alpha_I}dr+c.c.~,
}
where here  the $\chi_{m,\mu_i}^{(j),\pm}$ carry a non-trivial wave function dependence at small $m_Wr$. 

\begin{figure}
\begin{center}
\includegraphics[scale=1,clip,trim=0.9cm 20cm 8cm 5cm]{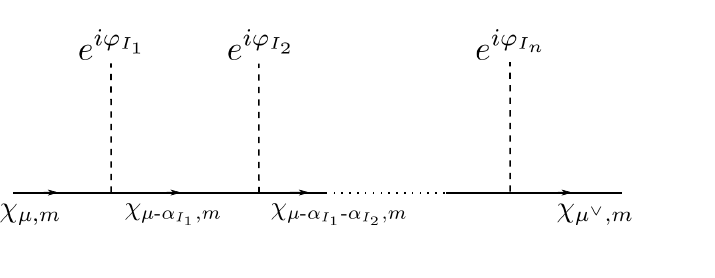}
\caption{This figure illustrates the Feynman diagram that gives rise to the interaction \eqref{dyonint1}. Here $\{\alpha_I\}$ are a set of simple positive roots.   \label{fig:FeynmanDiagram} }
\end{center}
\end{figure}

We know from the explicit zero-mode solutions that only  $\chi_{m,\mu_{top}}^{(j),-}$ and $\chi_{m,\mu_{bot}}^{(j),+}$ correspond to long range fermion modes. Thus, we should integrate out the other fermion modes. Doing so leads to the interaction \eqref{dyonint1} as shown in the Feynman diagram in Figure \ref{fig:FeynmanDiagram}. 

This tree level interaction is reflected in the structure of the fermion zero-modes.  We can think of the  virtual excitations of the fermion zero-modes that give rise to the above diagram as imprinted in the sub-leading structure of the zero-mode $\psi_{R_A}^{(j)}$ that vanishes at $r\to \infty$:
\eq{
\lim_{r\to \infty}r\psi_{R_A}^{(j)}=\psi_{\mu_{top}}^{(j)}+\psi_{\mu_{bot}}^{(j)}+\underbrace{\sum_{\mu_a\neq\mu_{top/bot}}\frac{1}{r^{\delta_a}}\psi^{(j)}_{\mu_a}}_{virtual}\quad \delta_a>0~.
} 

Here, the Feynman diagram clearly shows that the interaction above for fixed $R_A$ and $j$ couples $\psi_{\mu_{top}}^{(j)}\psi_{\mu_{bot}}^{(j)}$ to $e^{ic^I_\mu \varphi_I}$ where $c^I_\mu$ counts the number of $E_I^-$ required to map $\mu \to \mu^\ast$ 
\eq{
\prod_I (E_I^-)^{c^I_\mu}v_{\mu}=v_{\mu^\ast}~,
}
and is thus manifestly positive by the highest weight construction. 
Thus, we generally find interactions of the form 
\eq{
S_{int}=\sum_{m,\mu \in S_{R_A}}\int d^2x\,g_\mu(r)e^{ic^I_\mu\varphi_I} \chi_{\mu,m}^\dagger\chi_{\mu^\ast,m}+c.c. 
}
where here $\chi_{\mu,m}$ are the  long range fields and $g_\mu(r)$ is a function that encodes the overlap of the 4D fermion modes and the dyon degree of freedom.

We now need to worry about the fact that higher angular momentum modes fermion modes vanish at the origin as we previously alluded to. Recall that the dyon degree of freedom is exponentially localized to the monopole core while the long range fermion modes are generically repulsed:
\eq{
 a_I(r)\sim  e^{-\Delta_I m_W r}+...\quad, \quad f_{\mu_{top},-},f_{\mu_{bot},+}(r)\sim (m_Wr)^\ell+...~.
} 
However, even though the dyon degree of freedom becomes infinitely localized to the monopole core the wavefunction overlap does not vanish in the limit of $m_W\to \infty$. 

To show this, it is sufficient to note that the integral of the coupling function 
\eq{
\int g_\mu(r)dr&=\int \prod_I\big( a_I(r)\big)^{c^I_{\mu_{top}}}f_{\mu_{top},-}^\dagger(r)f_{\mu_{bot},+}(r) \frac{dr}{r}~,
%&=\int m_W a_I(m_W r) f_{\mu_{top},-}(m_Wr)f_{\mu_{bot},+}(m_Wr) dr
}
can be recast as an integral over the dimensionless quantity $x=m_Wr$ and therefore is independent of the limit $m_W\to \infty$. Further, since the function can be well approximated by the leading behavior, we can see that the  integral is non-zero\footnote{
In the case where $\ell=0$ the coupling derived here requires regulation. The reason why the regulated integral makes sense here is that we could alternatively have absorbed the coupling to $\varphi_Ic^I_\mu$ into $\psi_\mu,\psi_{\mu^\ast}$ so that the singular term in question is simply the coupling to the background gauge field. Then since the $\psi_\mu$ are assumed to solve the Dirac equation, we see that the form of the $f_\mu, a_I$ must be so that seemingly singular term contributes no singularity. This is an alternative derivation of the coupling \eqref{dyonthetacoupling}. 
}
\eq{
\int g_\mu(r)dr\simeq \int m_W e^{-c_\mu^I\Delta_I m_W r} (m_Wr)^{2\ell}\frac{dr}{r}\approx \frac{\Gamma(2\ell)}{(c^I_\mu \Delta_I)^{2\ell}}\neq 0~.
}
 
Since this integral is independent of $m_W$, non-zero, and the integrand is exponentially peaked near $m_Wr=0$, we see that in the IR limit where $m_W\to\infty$ we can approximate  
$g_\mu(r)= \delta(r)$. Here the normalization of $g_\mu(r)$ can be fixed by the asymptotic form of the zero-mode spin-$j$ zero-mode relating $\chi_{\mu_{top},m}^{(j)-}$ and $\chi_{\mu_{bot},m}^{(j)+}$.

\subsubsection{Fermion-Dyon Coupling}

Now we will explicitly determine the couplings $c^I_\mu$. As we discussed, $c^I_\mu$ is fixed by gauge invariance and can be computed explicitly in terms of  $SU(N)$ representation theory. %The casual reader may safely skip this section. 
 
Let us restrict to an irreducible $SU(2)_T$ component $\psi_{R_A}$ of $\psi_R$. 
Generically, only a subset of the positive weights have corresponding fermion zero-modes: those that belong to the subspace $ S_{R_A}$. So, for each weight 
 $\mu\in  S_{R_A}$, there exists a fermion zero-mode of spin-$j=\langle \mu,T_3\rangle-\half$. The coupling $c^I_\mu$ for such a weight is defined to be the number of lowering operators necessary to take $v_{\mu}\to v_{\mu^\ast}$ where again $\mu^\ast$ is the charge conjugate of $\mu$ with respect to $SU(2)_T$\footnote{More precisely, $\mu^\ast=w(\mu)$ where $w$ is the unique element of the Weyl group of $SU(2)_T$. This is the ``charge conjugate'' because it maps $q_\mu=\langle T_3,\mu\rangle\to q_{\mu^\ast}=-q_\mu$. }.

To facilitate this calculation, we will use Dynkin indices for representations of $SU(N)$. Here a representation $R_\Lambda$ of highest weight $\Lambda$ is denoted by a vector:
\eq{
\Lambda=\sum_{I=1}^{N-1} n_I \lambda^I\longmapsto [n_1,n_2,...,n_{N-1}]\quad, \quad (\lambda^I,\lambda^J)=C^{IJ}~, 
}
where the $\lambda^I$ are a basis of positive simple weights of $SU(N)$,  $C^{IJ}$ is the inverse Cartan matrix, and $(~,~)$ is the Killing form on $\fsu(N)$. For $SU(N)$ the raising and lowering operators $E_{I}^\pm$ act on a vector $v_\mu$ by shifting the weights as
\eq{
E^\pm_I v_\mu&=v_{\mu\pm \alpha_I}~,\\
\mu&=[n_1,...,n_{I-1},n_I,n_{I+1},...,n_{N-1}]~,\\
\mu\pm \alpha_I&=[n_1,...,n_{I-1}\mp 1,n_I\pm 2,n_{I+1}\mp 1,...,n_{N-1}]~.
}
%for $\alpha_I$ a positive simple root. 
All of the$SU(N)$ representation $R_\Lambda$  are expressible by successive action of the $E^-_{I}$ on the vector of highest weight $v_\Lambda$. This is the standard construction of highest weight representations. 

It will also be convenient to introduce a notation for the vectors in the dual space (co-root lattice):
\eq{
\widetilde\Lambda=[\tilde{n}_1,...,\tilde{n}_{N-1}]^\vee ~, 
}
where $\langle \widetilde\Lambda,\Lambda\rangle=(\widetilde\Lambda^\vee,\Lambda)$ for 
\eq{
\widetilde\Lambda^\vee=\sum_{I,J} \tilde{n}_I C_{IJ}\lambda^J~.
}
In this notation, we can also encode $T_3,I_3,$ and $\gamma_m$ as the vectors
\eq{
T_3=[t_1,t_2,...,t_{N-1}]^\vee\quad, \quad I_3=[i_1,i_2,...,i_{N-1}]^\vee\quad, \quad \gamma_m=[\gamma_1,\gamma_2,...,\gamma_{N-1}]^\vee~,
}
which  are related by 
\eq{
\gamma_m=T_3-I_3=[t_1-i_1,t_2-i_2,...,t_{N-1}-i_{N-1}]^\vee~. 
}
The action of the above Lie algebra generators on a vector with weight 
 $\Lambda=[n_1,...,n_{N-1}]$ can easily be computed in this notation: for example 
\eq{
\langle T_3,\Lambda\rangle=\sum_I n_I t_I~. 
}

We can now easily compute the couplings to the different $W$-bosons. Given a weight $\Lambda=[n_1,...,n_{N-1}]$, the charge conjugate weight is given by
\eq{
\Lambda\to \Lambda^\ast=[-n_{N-1},...,-n_1]~.
}
The number of lowering operators required to map a positive weight $\Lambda\to \Lambda^\ast$ is then given by 
\eq{
&C^I_\Lambda=\left(C_{\Lambda}^1\,,\,C_{\Lambda}^2\,,\,C_{\Lambda}^3\,,\,...\,,\,C_{\Lambda}^{\left\lfloor\frac{N-1}{2}\right\rfloor}\,,\,C_{\Lambda}^{\left\lfloor\frac{N-1}{2}\right\rfloor+1}\,,\,...\,,\,C_{\Lambda}^{N-1}\right)~,
}
where $C_{\Lambda}^I$ is the number of $E^-_{I}$ required to go from $\Lambda\mapsto  \Lambda_\vee$  which are given explicitly by 
\eq{
&C_{\Lambda}^1=\sum_I n_I
\\&C_{\Lambda}^2=(n_1+n_{N-1})+2\sum_{I\neq 1,N-1} n_I
\\&C_{\Lambda}^3=(n_1+n_{N-1})+2(n_2+n_{N-2})+3\sum_{I\neq1,2,N-1,N-2}n_I
\\&\vdots
\\&C_{\Lambda}^{\left\lfloor\frac{N-1}{2}\right\rfloor}=\sum_{M}M(n_M+n_{N-M})
\\&C_{\Lambda}^{\left\lfloor\frac{N-1}{2}\right\rfloor+1}=\sum_{M\neq \left\lfloor\frac{N-1}{2}\right\rfloor}M(n_M+n_{N-M})+(n_{\left\lfloor\frac{N-1}{2}\right\rfloor}+n_{N-\left\lfloor\frac{N-1}{2}\right\rfloor})
\\&\vdots
\\&C_{\Lambda}^{N-1}=\sum_I n_I
~.
}

However, since the low energy effective theory generically has non-abelian gauge symmetry, the $E^-_I$ that are not along broken directions in the $SU(N)$ gauge group in some sense ``come for free'' due to coupling to long range gluons.  
 Thus, the coupling $c^I_\Lambda$ to the dynamical dyon degrees of freedom is simply the truncation of the $C^I_\Lambda$ to the set  of $I$ such that $i_I=0$.

We then find that the low energy effective theory of fermions and dyons is given by 
\eq{
S=\sum_{\mu%\in \Delta^+_{R_A}[T_3,I_3]
,m}&\int d^2x\Big[i \chi^\dagger_{\mu,m}\partial_-\chi_{\mu,m}+i \chi^\dagger_{\mu^\ast,m}\partial_+\chi_{\mu^\ast,m}+i \delta(r) e^{i c^\mu_I \varphi_I}\chi_{\mu,m}^\dagger\chi_{\mu^\ast,m}+c.c.\Big]\\
&+\sum_I \int dt~\half m_W\, \dot\varphi_I^2~.
}
We can interpret the dyon interaction term as a boundary term that imposes the boundary condition 
\eq{
\chi_{\mu^\ast,m}=e^{-ic^\mu_I \varphi_I}\chi_{\mu,m}\big{|}_{r=0}~.
}
We can now ``unfold'' our theory to recast our theory in a form similar the model studied in \cite{Polchinski:1984uw}. By this, we mean to map the theory with a boundary interaction on the half-plane to a theory with a localized interaction on the full plane. 
To do this, let us introduce the coordinate 
\eq{
x=\begin{cases}
-r&x<0\\
r&x>0
\end{cases}
}
and identify the single chiral fermion 
\eq{
\Psi_{\mu,m}(t-x)=\begin{cases}
\chi_{\mu,m}(t+x)&x<0\\
\chi_{\mu^\ast,m}(t-x)&x>0
\end{cases}
}
on the $(x,t)$-plane. We will also use the field redefinition
\eq{
\chi_{\mu,m}\big{|}_{r=0}\to e^{i\frac{c^\mu_I}{2} \varphi_I}\chi_{\mu,m}\big{|}_{r=0}\quad, \quad \chi_{\mu^\ast,m}\big{|}_{r=0}\to e^{-i\frac{c^\mu_I}{2} \varphi_I}\chi_{\mu^vee,m}\big{|}_{r=0}
}
to eliminates the above boundary condition/field discontinuity in exchange for the interaction term
\eq{\label{dyonthetacoupling}
\CL_{int}= \Theta(r) c^{(\mu)}_I \dot\varphi_I \Psi^\dagger_{\mu,m}\Psi_{\mu,m}~,
}
where here $\Theta(r)=\Theta_{heviside}(r)-\half$. The unfolded action is now 
\eq{\label{unfoldedaction}
S=\sum_{\mu\in S_R,m}\int d^2x\left[i\Psi^\dagger_{\mu,m}\Big(\partial_+- i \Theta(x) c^I_{\mu} \dot\varphi_I\Big)\Psi_{\mu,m}\right]+\sum_I\int dt\,\half m_W \dot\varphi_I^2~.
}

\subsection{Effective Boundary Condition}

We would now like to examine the dynamics of the effective theory above in order to determine the effective boundary conditions for the fermions. These boundary conditions can be found by simply integrating out the $\varphi_I$ degrees of freedom. 
The physics of the dyon-fermion system are very similar to the toy model considered in \cite{Polchinski:1984uw}.  Here we will follow a similar analysis to integrate out the $\varphi_I$ degrees of freedom.

Let us consider the theory described by the action in \eqref{unfoldedaction}. To facilitate the computations in this section we will introduce an index $A$ for the fermions $\Psi_A$ that encodes both $\mu\in S_R$ and $m=-j_\mu,...,j_\mu$. With this notation, the action can be written 
\eq{
S=\sum_{\mu\in S_R,m}\int d^2x\left[i\Psi^\dagger_{\mu,m}\Big(\partial_+- i \Theta(x) c^I_{\mu} \dot\varphi_I\Big)\Psi_{\mu,m}\right]+\sum_I\int dt\,\half m_W \dot\varphi_I^2~.
} 

This theory has a series of currents that correspond to a $U(1)$ rotation of each $\Psi_A$ individually:
\be
\Psi_B\to e^{i \phi \delta_{AB}}\Psi_B\quad\Rightarrow\quad J_A(x)=\Psi^\dagger_A(x)\Psi_A(x)~.\ee
The corresponding symmetries have ABJ-type anomalies and consequently their currents are not conserved. Since the coupling to $\varphi_I$ appears as a gauge coupling, the conservation equation for the current is given 
\eq{
\partial_+J_A=\dot\varphi_Ic^I_A \delta(x)~.
} 
The action is also manifestly shift invariant under $\varphi_I\to \varphi_I+\beta$ for any constant $\beta$. There is a corresponding conserved charge: 
\be
Q=%\Pi+\int dr \,q(r)\Psi^\dagger_A(r) \Psi_A(r)=
m_W \dot\varphi+\int dx\, \Theta(x) \sum_A J_{A}(x)~. \ee 
Together, the conservation equations satisfy:
\begin{align}\begin{split}
\partial_+J_A=-\frac{c^I_A}{m_W}\Pi_I\delta(x)\quad, \quad 
&\partial_t\Pi_I=\sum_A c^I_A\int dx \,\delta(x)J_A(x)~,
\end{split}\end{align}
where $\Pi_I$ is the conjugate momentum to $\varphi_I$.

Integrating the conservation equations  along a line of constant $x-t$ (incoming plane wave) produces
\begin{align}\begin{split}
&J_A^{(out)}=J_A^{(in)}-\frac{c^I_A}{m_W}\Pi_I\delta(x)~,\\
&\partial_t\Pi_I= \sum_A c^I_AJ_A^{(in)}-\frac{1}{m_W}\sum_A c^I_A c^J_A\Pi_{J}~. 
\end{split}\end{align}
which means that the incoming charge differs from the out-going charge by something proportional to the momentum of the periodic scalar.

%These equations imply that $\varphi_I$ excitations will tend to decay exponentially. This is apparent from the case without any incoming current:
%\eq{
%\partial_t \langle \Pi_I\rangle=-\sum_A\frac{c^I_Ac^J_A}{m_W}\langle \Pi_J\rangle~.
%}
%by using the fact that the eigenvalues of the matrix $c^I_Ac^J_A$ are positive definite\footnote{In general a matrix that is the . 
We can now integrate out the $\varphi_I$ in which we effectively set 
\eq{
\partial_t\langle \Pi_I\rangle=0\quad\Rightarrow\quad c^I_A\Pi_I=m_W J_A~. 
}
We want to solve this formula for $\Pi_I$ and plug back into the $J_A$ current conservation equation, but this is in general non-trivial since $c^I_A$ is a non-rectangular matrix. However, the physics of this equation is clear: since $c^I_A$ is the coupling between $\Pi_I,J_A$ any virtual excitation of the $\varphi_I$ will decay via the bulk current of charge of the associated gauge charge:
\eq{
S_I=\sum_A c^I_A\int \Theta(x)\dot\varphi_I J_A\quad \Rightarrow\quad 
\varphi_I\to J_I=\frac{1}{N_I}\sum_A c^I_A J_A~,
}
where here $N_I$ is a normalization constant that is fixed by gauge charge conservation.

Each $\varphi_I$ has charge 2 under the associated $U(1)_I$ gauge symmetry\footnote{This comes from our choice of convention $[H_I,E^\pm_I]=\pm 2E^\pm_I$.}. 
For each $\Psi_A$ there is an associated weight $\mu^{(A)}=[\mu_1^{(A)},...,\mu_{N-1}^{(A)}]$. The charge of the out-going $\Psi_A$ under $U(1)_I$ is then given by $-\mu^{(A)}_{N-I}$ and the total charge of the current is given by 
\eq{
N_I\times Q_I[ J_I]=-\sum_A c^I_A \mu^{(A)}_{N-I}~. 
}
Thus, we find that the normalization $N_I$ is given by 
\eq{
N_I=-\half\sum_A c^I_A \mu^{(A)}_{N-I}~.
}
Therefore, the current conservation equation is given by 
\eq{
J_A^{(out)}=J_A^{(in)}-\sum_{I,B}\frac{2}{\sum_D c^I_D\mu^{(D)}_{N-I}} c^I_A c^I_B j_B~.
}
Translating into the notation of section \ref{sec:2}, the corresponding effective boundary condition is specified by 
\eq{
\bar{J}_A=\CR_{AB}J_B\quad, \quad \CR_{AB}=\delta_{AB}-\sum_{I}\frac{2}{\sum_A c^I_A\mu^{(A)}_{N-I}} c^I_A c^I_B~.
}

Thus, when the dyon is excited by an incoming fermion, the dyonic excitation dissipates through all possible fermion number channels. This can be interpreted as soft, charged radiation made up of fermionic modes that carries away the charge and energy from the monopole. This radiation generically can carry fractional flavor charges since the flavor symmetry is not actually preserved in the quantum theory. This soft radiation is sometimes referred to in the literature as `` propagating pulses of vacuum polarization'' \cite{Polchinski:1984uw}. 

%It is often said following the discussion of \cite{Polchinski:1984uw} that this soft radiation can be thought of as propagating pulses of vacuum polarization.

\section{Preserved Symmetries}

\label{sec:5}

We would now like to determine what global symmetries are preserved by the boundary conditions specified by a choice of $\CR$. 
We would like to consider a collection of fermions that transform under representations $\{R_i\}$ in the presence of an $SU(N)$ monopole. 
%\eq{
%R=\bigoplus_i R_i~.
%}
%Such a theory has a global symmetry which is  given by the stabilizer subgroup of $R$ in $SU(r)$ where $r=dim[R]$. 

Let us consider a $U(1)$ symmetry  of the fermion fields that rotates each $\psi_{R_i}$ individually:
\eq{
\psi_{R_i}\to e^{i q_i \phi}\psi_{R_i}~.
}
The charges of the in-going/out-going low energy fermion modes define the vectors $q_A^{(in)},q_A^{(out)}$. % of charges of the in-going/out-going fermion modes. 
The symmetry is preserved iff the in-going charge is equal to the out-going charge:
\eq{\label{symcond}
q_A^{(in)}= \CR_{AB}\,{q}_B^{(out)}~. 
}
For symmetries that commute with the $SU(N)$ gauge symmetry $q_A^{(in)}=q_A^{(out)}=q_A$  
 and therefore the associated global symmetry is in fact preserved if $q_A$ is a unit eigenvector of the matrix $\CR_{AB}$.  
 
 More generally, we can consider IR symmetries that are a linear combination of UV gauge symmetry with UV global symmetries such as $B-L$ symmetry in the standard model. In this case $q_A^{(in)}\neq q_A^{(out)}$ and the condition that the symmetry is preserved by the boundary condition is \eqref{symcond}.

In general,  the boundary conditions specified by a matrix $\CR$ can preserve non-abelian symmetries which we can infer by studying the preserved abelian symmetries \cite{Smith:2020nuf}. Given a choice of $\CR$, one can construct an integer lattice $\Lambda_\CR$ that is generated by the charge vectors of the preserved abelian symmetries. The boundary conditions specified by $\CR$ then preserve a semi-simple non-abelian symmetry group $G$ when $\Lambda_\CR$ contains the weight lattice of $G$ as a sublattice $\Lambda_{wt}[G]\subset \Lambda_R$.

We would now like to determine under what conditions a global symmetry is violated by the monopole boundary conditions. 
Due to the UV physics encoded by the monopole boundary conditions it is clear that a global symmetry will be broken if the process 
\eq{\label{decayalpha}
\varphi_I\longrightarrow \frac{1}{N_I}\sum_A c^I_A J_A~,
}
violates the global symmetry.  

Let us consider a global symmetry $U(1)_q$ under which a fermion of representation $R$  has charge $q_R$. Since $\varphi$ is uncharged under $U(1)_q$, the violation of $U(1)_q$ is measured by the charge of the right hand side of \eqref{decayalpha}:
\eq{
\Delta Q_q^{(I)}=\frac{1}{N_I}\sum_A q_A^{(out)} c^I_A=\frac{q_R}{N_I} \sum_{\mu\in S_R} m_\mu c_\mu^I~,
}
where $m_\mu$ is the number of fermions with weight $\mu\in S_R$. Now, since  $c^I_\mu$ encodes the gauge charges of the operator $\chi_{\mu,m}^\dagger \chi_{\mu^\ast,m}$, we can compute $c^I_\mu$ as the difference in charge
\eq{
c^I_\mu=Q_I[\psi_\mu]-Q_I[\psi_{\mu^\ast}]~,
}
 under the corresponding $U(1)_I$ gauge symmetry generated by 
\eq{
h^I=\frac{1}{N}\diag(\underbrace{N-I,...,N-I}_{I\text{-times}},\underbrace{-I,...,-I}_{N-I\text{-times}})~.
}
Thus we find that the total charge violation is computed by 
\eq{
\Delta Q_q^{(I)}=\frac{1}{N_I}\sum_R\sum_{\mu\in S_R} q_R\left(Q_I[\psi_\mu]-Q_I[\psi_{\mu^\ast}]\right)m_\mu~, 
}
which proportional to the coefficient of the 2D ABJ-type anomaly of $U(1)_q$. Since the 2D theory is simply the low energy effective theory of the 4D $SU(N)$ theory, the 2D anomaly must match an ABJ anomaly of the full 4D theory. Therefore, if a global symmetry has no ABJ anomaly, then $\Delta Q_q^{(I)}=0$ and the scattering process will preserve the symmetry regardless of 't Hooft anomalies. 

This relation can be seen explicitly in the case where $\vI=0$. From \cite{Brennan:2021ucy}, we know that the multiplicity of fermions with weight $\mu\in S_R$ is given by 
\eq{
m_\mu=2\langle T_3,\mu\rangle~,
}
and additionally, $c^I_\mu$ can be computed by 
\eq{
c^I_\mu=\langle \rho_I,\mu\rangle\quad, \quad \rho_I=[1,2,3,...,\underbrace{I,...,I}_{\small{N-2I+1}},I-1,...,1]^\vee~.
}
We can now write 
\eq{
\sum_{\mu\in S_R} m_\mu c_\mu^I=2\sum_{\mu\in \Delta_R^+}\langle T_3,\mu\rangle \langle \rho_I,\mu\rangle=\sum_{\mu \in \Delta_R}\langle T_3,\mu\rangle \langle \rho_I,\mu\rangle~,%\,\sign(\langle \mu,T_3\rangle)~,
}
where here we used the fact that $c^I_\mu\geq 0$ if $\langle T_3,\mu\rangle>0$ which follows form the fact that $\langle \mu,\rho_I\rangle=c^I_\mu$, which is the number of $E^-_I$ required to go from the weight $\mu\to \mu^\ast$, is manifestly positive in the construction of highest weight representations. 

We can now rewrite the charge violation coming from $\Psi_{R}$ as 
\eq{
\Delta Q_q^{(I)}= \frac{q_R}{N_I}\times \Tr_{R}[T_3\rho_I]~, 
}
where $\Tr_{R}$ is the trace in the $R$ representation and 
\eq{
\rho_I=\diag(\underbrace{1,...,1}_{I},0,...,0,\underbrace{-1,...,-1}_{I})~.
}
The contribution to global charge violation can then be further simplified to 
\eq{
\Delta Q_q^{(I)}=\frac{q_R I_2(R)}{N_I}\times \Tr_{f}[T_3\rho_I]~, 
}
where the trace is now taken in the fundamental representation. The prefactor $q_R I_2(R)$ matches the anomaly coefficient of an ABJ-anomaly between $U(1)_q$ and $SU(N)$ which indicates that the charge violation arises from such anomalies. 
 
Therefore, we see that the fermion boundary conditions on the monopole explicitly violate global symmetries that have ABJ anomalies.  

\section{Extended Examples}

\label{sec:6}

Here we will compute the scattering/effective boundary conditions for a variety of examples in order to explicitly demonstrate the above formalism. 

\subsection{$SU(2)\to U(1)$ with Arbitrary Representations}

Let us take $G=SU(2)$ gauge theory with a collection of fermions of representations of spins $\{j_I\}$. Let us consider the minimal monopole which has 
\eq{
\gamma_m=2T_3=\diag(1,-1)\quad, \quad \vI=0~,
}
and $\Phi_\infty=v T_3$ for $v\in \IR$. This Higgs field breaks the gauge symmetry to $U(1)$ generated by $\gamma_m$. 
Here we find that the highest weight of the representation of spin-$j_I$ is
\eq{
\Lambda_I=[2j_I]\quad, \quad T_3=\left[\half\right]^\vee~.
}
Each fermion then has 
\eq{
\sum_{i=1}^{\lceil j_I\rceil} 2(j_I+1-i)=\begin{cases}
j_I(j_I+1)&j_I\in \IZ\\
\left(j_I+\half\right)^2&j_I\in \IZ+\half
\end{cases}
}
total zero-modes. If we index the fermion zero-modes by $\Psi_{i,a}$ where \eq{i=1,...,\lceil j_I\rceil~,~a=1,...,i~, }
each of which has out-going charge 
\eq{
Q[\Psi_{i,a}]=-2(j_I+1-i)
}
then we find that the dyon decays 
\eq{
\varphi\to-\frac{1}{\sum_I\sum_{i=0}^{\lfloor j_I\rfloor} 2(j_I+1-i)^2} \sum_{i,a}\bar\Psi_{i,a}\Psi_{i,a}~,
}
so that the charges are given 
\eq{
 -\frac{1}{\sum_I\sum_{i=0}^{\lfloor j_I\rfloor}2 (j_I+1-i)^2} \sum_{i,a}Q[\Psi_{i,a}]=1~.
}
Relabeling $(i,a)$ by a single index $A$, we see that the boundary condition are given by 
\eq{
J_{A}\to \CR_{AB}\bar{J}_{B}\quad, \quad \CR_{AB}=\begin{cases}
1-\frac{3}{\sum_Ij_I(j_I+1)(2j_I+1) }&A=B\\
\frac{3}{\sum_Ij_I(j_I+1)(2j_I+1) }&A\neq B
\end{cases}
%\mathds{1}-\frac{2}{\sum_I\sum_{i=0}^{\lfloor j_I\rfloor} (2j_I-2i)^2} =\mathds{1}-
%\frac{3}{\sum_Ij_I(j_I+1)(2j_I+1) }~.
}
As an example, for the case of $N_f$ fundamental fermions  (where $\{j_I\}_{I=1}^{N_f}$ with $j_I=\half$), we find 
\eq{
\CR_{AB}=
\begin{cases}1-\frac{2}{N_f}&A=B\\
-\frac{2}{N_f}&A\neq B
\end{cases}~,
}
matching the known boundary condition of \cite{Callan:1982ac,Callan:1982ah,Callan:1982au,Rubakov:1982fp,Maldacena:1995pq,Affleck:1993np}.

\subsection{$SU(N)\to U(N-1)$ with $N_f$ Fundamental Fermions}

Let us take $G=SU(N)$ gauge theory with $N_f$ fundamental fermions. Let us consider a class of monopoles that breaks $SU(N)\to \frac{SU(N-1)\times U(1)}{\IZ_{N-1}}=U(N-1)$. 
These correspond to $SU(N)$ fermions where $\vT$ is the maximal embedding and $\vI$ corresponds to the $(N-1)$-dimensional representation of $SU(2)_I$. Explicitly:
\eq{
T_3&=\half\diag(N-1,N-3,...,-N+3,-N+1)~,\\
I_3&=\half\diag(N-2,N-4,...,-N+2,0)~,\\
\gamma_m&=\diag(1,1,1,...,-N+1)~.
}
Additionally, $\Phi_\infty=v\,\gamma_m$ for $v\in \IR$. This Higgs vev spontaneously breaks $\fsu(N)$ to $\fsu(N-1)\times \fu(1)$ where the $\fu(1)$ is generated by $\gamma_m$. Additionally since $i_I\neq 0$ except for $I=N-1$, only $\varphi_{N-1}$ is dynamical.

In bracket notation, these are given by 
\eq{
T_3=&[t_I]^\vee\quad, \quad t_I=\half I(N-I)~,\\
I_3=&[i_I]^\vee\quad, \quad i_I=\half I(N-1-I)~,\\
\gamma_m=&[I]^\vee~.
}
Now, highest weight of the fundamental representation is given by 
\eq{
\lambda_1=[1,0,...,0]
}
The weight space is given by 
\eq{
\lambda_i=[\underbrace{0,...,0}_{i-2},-1,1,0,...,0]\quad, \quad i=2,...,N-1~,
}
along with $\Lambda_{N}=[0,...,0,-1]$. 

We can now compute the states that have non-trivial zero-modes. To do this, we must first note that 
\eq{
\lambda_i^\ast=\lambda_{N-i+1}~.
}
Then, using the fact that 
\eq{
\sign(\langle \gamma_m,\lambda_i\rangle)\times\sign(\langle\gamma_m,\lambda^\ast_i\rangle)&>0\quad i\neq 1,N~,\\
\sign(\langle \gamma_m\lambda_1\rangle)\times\sign(\langle \gamma_m,\lambda_N\rangle)&<0~,
} 
the are only zero-modes associated to the pair of dual weights $\lambda_1,\lambda_N$ which form a spin $j=\frac{N-2}{2}$  multiplet. 

For $N_f$-fundamental fermions, there are $N_T=N_f\times (2j+1)=N_f(N-1)$ total low energy modes. The coupling  matrix is then given by a $N_T$-length vector $c_i=1$ $\forall i$ where here $i$ indexes all $(N-1)N_f$ low energy fermion currents. 
For this configuration, the ingoing/outgoing fermion has $U(1)$ charge 
\eq{
Q[\Lambda_f]=1
\quad, \quad Q[\Lambda_f^\ast]=-N+1~,
}
and the $U(1)$ charge of $\varphi_{N-1}$ is 
\eq{
Q[\varphi_{N-1}]=(N-1)-(-1)=N~.
}
This allows us to compute the decay  
\eq{
\varphi_{N-1}\to  
-\frac{N}{N_f(N-1)^2}\sum_B \bJ_B~.
}
This leads to the general scattering 
\eq{
J_A \to \bJ_{A}+\varphi_{N-1}\to \bJ_{A}-\frac{N}{N_f(N-1)^2}\sum_B \bJ_B~,
} 
which corresponds to the boundary condition that is specified by the $\CR$ matrix:
\eq{
\CR_{AB}=\begin{cases}
1-\frac{N}{N_f(N-1)^2}&A=B\\
-\frac{N}{N_f(N-1)^2}&A\neq B
\end{cases}
}

\subsection{ $SU(N)\to \frac{SU(N-M)\times SU(M)\times U(1)}{\IZ_{N-M}\times \IZ_M}$ with $N_f$ Fundamental Fermions}

Now let us consider the case of $SU(N)$ gauge theory with $N_f$ fundamental fermions in the presence of a monopole that breaks $SU(N)\to \frac{SU(N-M)\times SU(M)\times U(1)}{\IZ_{N-M}\times \IZ_M}$. Here we will take $M<N/2$.  This is realized by a $\vT$ which again is the maximal embedding and $\vI$ corresponds to the direct sum of a $(N-M)$- and $M$-dimensional embedding. Explicitly:
\eq{
T_3&=\half\diag(N-1,N-3,...,-N+3,-N+1)~,\\
I_3&=\diag\left(\frac{N-M-1}{2},\frac{N-M-3}{2},...,\frac{-N+M+1}{2}\right)\\
&\qquad\oplus\diag\left(\frac{M-1}{2},\frac{M-3}{2},...,\frac{-M+3,-M+1}{2}\right)~,\\
\gamma_m&=\diag(\underbrace{M,...,M}_{N-M\text{ times}})\oplus \diag(\underbrace{M-N,..,M-N}_{M\text{ times}})~.
}
Here we use the notation $\oplus$ to denote concatenation. 
Additionally, $\Phi_\infty=v\,\gamma_m$ for $v\in \IR$. This Higgs vev spontaneously breaks the generators $\fsu(N)$ to $\fsu(N-M)\oplus \fsu(M)\oplus \fu(1)$ where the $\fu(1)$ is generated by $\gamma_m$. 

In bracket notation, these are given by 
\eq{
T_3=&[t_I]^\vee\quad, \quad t_I=\half I(N-I)~,\\
I_3=&\left[\frac{I(N-M-I)}{2}\right]^\vee\oplus [0]\oplus \left[\frac{I(M-I)}{2}\right]^\vee~,\\
\gamma_m=&[I M]^\vee\oplus[(N-M)(M-I+1)]^\vee~.
}
Note that since $i_{N-M}=0$ that only the $\varphi_{N-M}$ will be dynamical. 

Now we can explicitly show that the top $M$-states have non-trivial zero-modes: they correspond to the positive weights
\eq{
\lambda_1=[1,0,...,0]\quad, \quad \lambda_i=[\underbrace{0,...,0}_{i-2},-1,1,0,...0]~.
}
 These states, have spins 
\eq{
j_i=\frac{i(N-i)-1}{2}\quad, \quad i=1,..,M~,
}
and have in-going and out-going electric charge:
\eq{
Q^{(in)}=M\quad, \quad Q^{(out)}=M-N~.
}
Additionally, $Q[\varphi_{N-M}]=N$. 

Now, using the fact that the coupling to $\varphi_{N-M}$ of all fermion zero-modes is exactly 1, and the fact that there are a total of 
\eq{
N_{T}=N_f\times \sum_{I=1}^M2\frac{I(N-I)-1}{2}=N_f\frac{M(M-1)(3N-2M-1)}{6}~,
}
fermion zero-modes which we index by a general label $A$, 
we find that the matrix $c^{N-M}_A$ again reduces to a $N_f\frac{M(M-1)(3N-2M-1)}{6}$-dimensional vector 
\eq{
c^{N-M}_A=1\quad, \quad \forall A~. 
}
Again, we can compute the radiation of the monopole d.o.f.
\eq{
\varphi_{N-M}\to%& \frac{1}{N}\sum_Ac_A^{N-M} \bJ_{A}=\frac{1}{Q[\psi_A]\times N_T}\sum_Ac_A^{N-M} \bJ_{A}\\
&=-\frac{6N}{(N-M)N_fM(M-1)(3N-2M-1)}\sum_Ac_A^{N-M} \bJ_{A}~. 
}
Again this leads to the boundary condtions specified by 
%\eq{
%J_A\to \CR_{AB}\bJ_B~,
%}
%where 
\eq{
\CR_{AB}=\begin{cases}
1-\frac{6N}{(N-M)N_fM(M-1)(3N-2M-1)}&A=B\\
-\frac{6N}{(N-M)N_fM(M-1)(3N-2M-1)}&A\neq B
\end{cases}
}
which is again a rational matrix. 

\subsection{$SU(N)\to U(N-1)$ with $N_f$ Adjoint Fermions}

Let us now consider the case of an $SU(N)$ monopole that breaks  $SU(N)\to \frac{SU(N-1)\times U(1)}{\IZ_{N-1}}$ in a theory with $N_f$ adjoint fermions. Here, the positive weights are given by $n$-tuples of positive adjacent simple roots:
\eq{
\alpha_I\quad, \quad \alpha_I+\alpha_{I+1}\quad, \quad\alpha_I+\alpha_{I+1}+\alpha_{I+2}\quad...
}
Together with the negative $n$-tuples of adjacent simple roots and $(N-1)$ zero-weights (parameterizing the Cartan algebra) the positive weights generate the $(N^2-1)$-dimensional adjoint representation. 
%and the negative weights are simply $(-1)$ times the positive weights.
Here we will use the explicit parametrization of simple roots:
\eq{
\alpha_1=[2,-1,0,...,0]\quad, \quad \alpha_I=[\underbrace{0,..,0}_{I-2},-1,2,-1,0,...,0]\quad, \quad \alpha_{N-1}=[0,...,0,-1,2]~.
}
Their complex conjugate weights are given by
\eq{\label{weightduality}
\alpha_I^\ast=-\alpha_{N-I}~. 
}
Now by noting that $\gamma_m=[I]^\vee$ and 
\eq{
 \langle \gamma_m,\alpha_I\rangle=N \,\delta_{I,N-1}~,
}
we see that  the low energy modes can only come from fermions with weights that contain $\alpha_{N-1}$. However, due to the duality of weights \eqref{weightduality}, we see that the only pair of fermions that contributes to the low energy theory is the highest/lowest weight pair:
\eq{
\langle \gamma_m,\mu\rangle\times \langle \gamma_m,\mu^\ast\rangle<0\quad\Rightarrow\quad \mu=\mu_{top}:=\sum_{I=1}^{N-1}\alpha_I~,
}
which couples to the dyon with weight $c^{N-1}_{\mu_{top}}=2$. Using the fact that $T_3=\half[I(N-I)]^\vee$ and hence $\langle T_3,\alpha_I\rangle=1$, the  associated low energy modes form a spin-$j=\langle T_3,\mu_{top}\rangle-\half=N-\frac{3}{2}$ multiplet. 
%Thus, there are $N_f(N-1)$ low energy modes with the corresponding positive weights:
%\eq{
%\mu_i=\sum_{j=1}^i \alpha_{N-j}\quad, \quad i=1,...,N-1~,
%}
%each solution of which belongs to a spin 
%$j_i=\frac{2i-1}{2}$-multiplet since $T_3=\half[I(N-I)]^\vee$ and hence $\langle T_3,\alpha_I\rangle=1$. 
Thus, there are a total  
\eq{
N_T=N_f\left(2N-2\right)~,
}
low energy fermions  which we index by the generalized index $A$.

We now find that the decay of $\varphi_{N-1}$ is described by 
\eq{
\varphi_{N-1}\to&% \frac{1}{\CN}\sum_{A}N\bJ_{A} 
-\frac{1}{2N_f(N-1)}\sum_A\bJ_A
}
This again gives rise to the boundary conditions 
\eq{
J_A\mapsto \CR_{AB}\bJ_B~,
}
where $\CR_{AB}$ is given by 
\eq{
\CR_{AB}=\begin{cases}
1-\frac{1}{N_f(N-1)}&A=B\\
-\frac{1}{N_f(1N-1)}&A\neq B 
\end{cases}
}

\subsubsection{Adjoint and Fundamental Matter}

Now we can put together the two sets of results and consider the case where there are $N_f$-fundamental fermions and $N_{ad}$-adjoint fermions. Again only $\varphi_{N-1}$ is dynamical, and its decay is described by
\eq{
\varphi_{N-1}\to \frac{1}{\CN}\sum_{A}n_A \bJ_{A}\quad, \quad n_A=\begin{cases}
1&\Psi_A \text{ from fundamental fermion}\\
2& \Psi_A \text{ from fundamental fermion}
\end{cases}
}
where the coefficient $\CN$ is given explicitly by 
\eq{
\frac{1}{\CN}& =\frac{N}{N_{T,f} \times Q[\psi_f]+2\times N_{T,ad}\times Q[\psi_{ad}]}\\
&=-\frac{N}{N_f(N-1)\times (N-1)+N\times 2N_{ad}(N-1)}
}
This leads to  boundary conditions described by the matrix
\eq{
\CR_{AB}=\begin{cases}
1-\frac{N}{N_f(N-1)^2+ 4N_{ad}N(N-1)} & A=B~\text{fund.}\\
1-\frac{4N}{N_f(N-1)^2+ 4N_{ad}N(N-1)}&A=B~\text{adjoint}\\
-\frac{N}{N_f(N-1)^2+ 4N_{ad}N(N-1)}&A\neq B,~A,B~\text{fund}\\
-\frac{4N}{N_f(N-1)^2+ 4N_{ad}N(N-1)}&A \neq B ~A,B~\text{adjoint}\\
-\frac{2N}{N_f(N-1)^2+ 4N_{ad}N(N-1)}&A ~\text{adjoint},~B~\text{fund or fund$\leftrightarrow$adjoint} \\
\end{cases}
} 
\subsection{$SU(5)$ GUT Monopoles}

Now let us consider the case of the spherically symmetric monopoles in the $SU(5)$ Georgi-Glashow GUT theory. In this model, the Higgs vev is given explicitly by 
\eq{
\Phi_\infty=v\,\diag(2,2,2,-3,-3)~,
}
 for $v\in \IR$ which breaks $SU(5)\to \frac{SU(3)\times SU(2)\times U(1)}{\IZ_6}$.  Here the $U(1)$ factor is referred to as \emph{hypercharge} and is generated by $Y=\diag(2,2,2,-3,-3)$. 

In this model, there is a $\IZ_6$ clafssification of monopoles that are compatible with this breaking reflecting the $\IZ_6$ quotient in the IR gauge group \cite{Gardner:1984zd}. %Reflecting the $\IZ_6$ quotient, there are 6 such monopoles. 
However, as shown in \cite{Gardner:1983uu}, only the charge 1,2,3 and 6 are stable, spherically symmetric monopoles. Thus, for brevity we will only consider the charge 1,2,3 monopoles. 

In the case of a single generation as we will consider here, the matter content of the $SU(5)$ gauge theory is a fermion in the $\bar{5}$- and the $10$-representations. These have weight spaces 
\eq{
\bar{5}:~\{\mu_i\}=&[-1,0,0,0]_1~,~[1,-1,0,0]_2~,~[0,1,-1,0]_3~,~[0,0,1,-1]_4~,~[0,0,0,1]_5~,\\
10:~\{\lambda_a\}=&[0,1,0,0]_1~,~[1,-1,1,0]_2~,~[-1,0,1,0]_3~,~[1,0,-1,1]_4~,~[-1,1,-1,1]_5~,\\&
[1,0,0,-1]_6~,~[-1,1,0,-1]_7~,~[0,-1,0,1]_8~,~[0,-1,1,-1]_9~,~[0,0,-1,0]_{10}~.
}
In terms of standard model fermions, the above weights correspond to 
\eq{
\bar{5}=&(d^c_1,d^c_2,d^c_3,\bar{e},\bar{\nu})~,\\
10=&(u_3^c,u_2^c,u_1^c,u_1,u_2,d_1,d_2,u_3,d_3,e^c)~.
}
Importantly, the  monopoles we consider here have an additional term to the Higgs vev as described in footnote \ref{footnotevev}: $\phi_\infty\neq 0$. This is chosen so that $\Phi_\infty$ always breaks $SU(5)$ to the standard model regardless of the magnetic charge. \\

\noindent\textbf{\underline{Charge 1 Monopole}}\\

\noindent Let us first consider the charge 1 monopole. This is the simple embedding of the $SU(2)$ monopole into the $SU(5)$ gauge group with 
\eq{
\gamma_m=2T_3=\diag(0,0,1,-1,0)=[0,0,1,0]^\vee~. 
}
Since $\gamma_m=2T_3$ and  $\vI=0$, we see that any fermion with positive charge will be in-going and will be paired to an out-going fermion with negative charge.  
After a simple computation, one can show that the pairs are given by
\begin{center}
\begin{tabular}{c|c|c|c}
in-going&out-going&$\varphi$-coupling&$j$\\
\hline\hline
$e$&$d_3^c$&$c_1=1$&0\\\hline
$d_3$&$e^c$&$c_1=1$&0\\\hline
$u_1^c$&$u_2$&$c_1=1$&0\\\hline
$u_2^c$&$u_1$&$c_1=1$&0
%$u_3^c$&$e^c$&$c_1,c_2=1$&$\half$
\end{tabular}
\end{center}
%\eq{
%(\mu_4,\mu_3)=(e,d_3^c)~, \quad (\lambda_9,\lambda_{10})=(d_3,e^c)~, \quad (\lambda_3,\lambda_5)=(u_1^c,u_2)~, \quad (\lambda_2,\lambda_4)=(u_2^c,u_1)~.
%}
Due to the fact that the fermions all couple to the dyon  with equal weight, we find that the decay of the single dyon mode is given by 
\eq{
\varphi\to \frac{1}{\CN}(d_3^c+e^c+u_2+u_1)~,
}
where here for notational simplicity we refer to the current by its component fermion field. 
$\CN$ can then be fixed by hypercharge conservation so that the decay of $\varphi$ is given by 
\eq{
\varphi\to -\frac{1}{2}(d_3^c+e^c+u_2+u_1)~,
}
so that the boundary condition is described by 
\eq{
\CR_{AB}=\half\left(\begin{array}{cccc}1&-1&-1&-1\\
-1&1&-1&-1\\
-1&-1&1&-1\\
-1&-1&-1&1
\end{array}\right)~,
}
which reproduces the famous Callan-Rubakov effect. \\

\noindent\textbf{\underline{Charge 2 Monopole}}\\

\noindent Now let us consider the charge 2 monopole. This is again a simple embedding of the $SU(2)$ monopole where $\vT$ decomposes as the $2\oplus 2\oplus 1$ embedding. This is usually written in the explicit basis 
\eq{
\gamma_m=2T_3=\diag(1,1,0,-1,-1)=[1,2,2,1]^\vee~. 
}
Again since $\vI=0$ the charges of the fermions under $T_3$ determine the spectrum of low energy fermions. 

The low energy fermion modes are paired up as:
\begin{center}
\begin{tabular}{c|c|c|c}
in-going&out-going&$\varphi_I$-coupling&$j$\\
\hline\hline
$e$&$d_1^c$&$c_1=1$&0\\\hline
$u_1^c$&$d_3$&$c_1=1$&0\\\hline
$\nu$&$d_2^c$&$c_2=1$&0\\\hline
$u_2^c$&$u_3$&$c_2=1$&0\\\hline
$u_3^c$&$e^c$&$c_1,c_2=1$&$\half$
\end{tabular}
\end{center}
By matching the charges of $\varphi_I$ and the out-going fermions we find the decay processes
\eq{
\varphi_1\to -\frac{1}{3}\left(d_1^c+d_3+2\times e^c\right)~,\\
\varphi_2\to -\frac{1}{3}\left(d_2^c+u_3+2\times e^c\right)~,
}
where again we refer to the currents by their component fermion fields and the multiplicity refers to the spin multiplicity.
If we write 
\eq{
J_A&=(e,u_1^c,u_2^c,u^c_{3,+},u^c_{3,-})~,\\
\bJ_{A}&=(d_1^c,d_3,d_2^c,u_3,e^c_+,e^c_-)~,
}
where $e^c_\pm,u^c_{3,\pm}$ are the $\pm$ states of the $j=\half$ angular momentum states, 
then we find that the boundary condition  
%\eq{
%J_A\to \CR_{AB}\bJ_B~,
%}
is specified by the matrix 
\eq{
\CR_{AB}=\left(\begin{array}{cccccc}
\frac{2}{3}&-\frac{1}{3}&0&0&-\frac{1}{3}&-\frac{1}{3}\\
-\frac{1}{3}&\frac{2}{3}&0&0&-\frac{1}{3}&-\frac{1}{3}\\
0&0&\frac{2}{3}&-\frac{1}{3}&-\frac{1}{3}&-\frac{1}{3}\\
0&0&-\frac{1}{3}&\frac{2}{3}&-\frac{1}{3}&-\frac{1}{3}\\
-\frac{1}{3}&-\frac{1}{3}&-\frac{1}{3}&-\frac{1}{3}&\frac{1}{3}&-\frac{2}{3}\\
-\frac{1}{3}&-\frac{1}{3}&-\frac{1}{3}&-\frac{1}{3}&-\frac{2}{3}&\frac{1}{3}
\end{array}\right)
} 

\noindent\textbf{\underline{Charge 3 Monopole}}\\

\noindent Now let us consider the charge 3 monopole which is not a simply embedded $SU(2)$ monopole. This is specified by the data
\eq{
\gamma_m=\diag(1,1,1,-1,-2)~, \quad T_3=\half\diag(2,1,0,-1,-2)~, \quad I_3=\half\diag(1,0,-1,0,0)~,
}
which is written in bracket notation as 
\eq{
\gamma_m=[1,2,3,2]^\vee\quad, \quad T_3=\half[2,3,3,2]^\vee\quad, \quad I_3=\half[1,1,0,0]^\vee~.
}
For this background the $\varphi_2,\varphi_3$ are dynamical degrees of freedom that interact with the pairs:
\begin{center}
\begin{tabular}{c|c|c|c}
in-going&out-going&$\varphi_I$-coupling&$j$\\
\hline\hline
${e}$&${d_2^c}$&$c_2=1$&0\\\hline
${u_{3}^c+u_{1}^c}$ &${e^c}$&$c_2,c_3=1$&1\\\hline
${ \nu}$&${d_{1}^c+d_{3}^c}$&$c_3=1$&$\half$\\\hline
${u_{1}^c}$&${d_{2}}$&$c_3=1$&0\\\hline
${u_{2}^c}$&${d_{1}+d_{3}}$&$c_3=1$&$\half$\\\hline
\end{tabular}
\end{center}
Again, we can compute the decay
\eq{
\varphi_2&\to-\frac{1}{4}( d_2^c+3 \times e^c)~,\\
\varphi_3&\to-\frac{1}{5}\Big( 2\times d_1^c+d_2+2\times(d_{1}+d_{3})+3\times e^c\Big)~,
}
where again we denot the current by its component fermion. 
The boundary boundary condition is then specified by the matrix
\eq{
\CR_{AB}=\left(\begin{array}{ccccccccc}
3/4&-1/4&-1/4&-1/4&0&0&0&0&0\\
-1/4&11/20&-9/20&-9/20&-1/5&-1/5&-1/5&-1/5&-1/5\\
-1/4&-9/20&11/20&-9/20&-1/5&-1/5&-1/5&-1/5&-1/5\\
-1/4&-9/20&-9/20&11/20&-1/5&-1/5&-1/5&-1/5&-1/5\\
0&-1/5&-1/5&-1/5&4/5&-1/5&-1/5&-1/5&-1/5\\
0&-1/5&-1/5&-1/5&-1/5&4/5&-1/5&-1/5&-1/5\\
0&-1/5&-1/5&-1/5&-1/5&-1/5&4/5&-1/5&-1/5\\
0&-1/5&-1/5&-1/5&-1/5&-1/5&-1/5&4/5&-1/5\\
0&-1/5&-1/5&-1/5&-1/5&-1/5&-1/5&-1/5&4/5
\end{array}\right)~.
}

\subsubsection{$B-L$ Symmetry in $SU(5)$ GUT Theory}

Here we would like to show that the boundary conditions from the charge 1,2,3 monopoles in the $SU(5)$ GUT theory preserves $B-L$ symmetry. $B-L$ symmetry can be realized as 
\eq{
Q_{B-L}=Q_Y+Q_D\quad, \quad D=\begin{cases}
-3& {\rm on}~\bar{5}\\
1&{\rm on}~10
\end{cases}
}
Since we have constructed the boundary conditions to preserve hypercharge, we simply need to check if $U(1)_D$ is also preserved by the boundary conditions. Since $U(1)^{(D)}$ commutes with $SU(5)$, it will be preserved by the boundary conditions if the corresponding charge vector is an unit eigenvector of the $\CR$-matrix. 

For the charge 1 monopole, the $D$ charge vector is given by 
\eq{
v_{D}=(-3,1,1,1)~,
}
which is indeed a unit-eigenvector of $\CR_{AB}$. 

For the charge 2 monopole, the $D$ charge vector 
\eq{\label{symcons}
v_D=(-3,1,-3,1,1,1)~.
}
Again, this is a unit eigenvector of $\CR_{AB}$. 

For the charge 3 monopole the $D$ charge vector is given by 
\eq{
v_D=(-3,1,1,1,
1,-3,-3,1,1)~,
}
which is again a unit eigenvector of $\CR_{AB}$. A similar computation shows that $B-L$-symmetry is also preserved by the charge 6 monopole. Therefore, $B-L$ symmetry is indeed preserved by the charge monopoles in the $SU(5)$ GUT model which is what we expect from the fact that there is no ABJ-type anomaly for the $U(1)_{B-L}$. 

However, now we can consider baryon number symmetry which does have an ABJ-type anomaly. Here the in-going/out-going symmetries for baryon number are:\footnote{Note that the charges here are 3$\times$ the standard charges in the particle physics literature. }
\eq{
{\rm charge~ 1:}~v^{(in)}_B&=(0,1,-1,-1)\qquad\qquad~~, \quad v^{(out)}_B=(-1,0,1,1)~,\\
{\rm charge~ 2:}~v^{(in)}_B&=(0,-1,0,-1,-1,-1)~, \quad v^{(out)}_B=(-1,1,-1,1,0,0)~,\\
{\rm charge~ 3:}~v^{(in)}_B&=(0,-1,-1,-1,-1,
0,0,-1,-1)~,\\
v^{(out)}_B&=(-1,0,0,0,1,-1,-1,1,1)~.
}
As expected, none of these vectors satisfy the conservation equation \eqref{symcons}. 
Therefore, baryon number symmetry (and therefore lepton number symmetry) is not preserved by the monopole boundary conditions and hence can be violated in scattering processes involving charge 1, 2,  and 3 monopoles. Similar results hold for the charge 6 monopole. 

%However, the ABJ-type anomaly preserves a $\IZ_3\subset U(1)_{B}$ which has a mixed `t Hooft anomaly with the $\IZ_3$ center symmetry of the standard model. This center symmetry is activated by the monopole charge mod 2 (i.e. the charge 1 and charge 3 monopoles activate the center symmetry). As we conjectured in section \ref{sec:discrete} we find that the $\IZ_3\subset U(1)_B$ is only preserved in the case of the charge 2 monopole:
%\eq{
%{\rm charge~ ,31:}~v^{(in)}_B-\CR\cdot   v^{(out)}_B\neq 0 {\rm mod}_3
%{\rm charge~ 2:}~v^{(in)}_B&=(0,-1,0,-1,-1,-1)~, \quad v^{(out)}_B=(-1,1,-1,1,0,0)~,\\
%{\rm charge~ 3:}~v^{(in)}_B&=(0,-1,-1,-1,-1,
%0,0,-1,-1)
%}

\section*{Acknowledgements}

We would like to thank   J. Harvey and G. Satishchandran for discussions and Sungwoo Hong for comments on the draft. TDB is supported by the  Mafalda and Reinhard Oehme Postdoctoral Fellowship in the Enrico Fermi Institute at the University of Chicago and in part by DOE grant DE-SC0009924. 

\appendix 
\section{Solving the Dirac Equation}
\label{app:A}

In this appendix we review the different solutions of the Dirac operator that are discussed in the main body.

\subsection{The Abelian Monopole}

Here we would like to solve for the solutions to the Dirac equation coupled to an abelian monopole connection. Here we will take the connection to be 
\eq{
A=\gamma_m A_{Dirac}=\frac{\gamma_m}{2}(1-\cos\theta)d\phi~,
}
where $\gamma_m\in \IZ$ is the magnetic charge and $\sigma=\pm1$. 
The time independent Dirac operator can be written 
\eq{
\bar\sigma^\mu D_\mu=\left(\sigma^{\hat{r}}\partial_r+\frac{1}{r}\sigma^{\hat{\theta}}\partial_\theta+\frac{1}{r\sin\theta}\sigma^{\hat{\phi}}\partial_\phi+\frac{i\gamma_m}{2r\sin\theta}\sigma^{\hat{\phi}}(1-\cos(\theta))\right)~,
}
where we use the notation 
\begin{align}\begin{split}
&\sigma^{\hat{r}}=\sigma^3\cos\theta+(\sigma^1\cos\phi+\sigma^2\sin\phi)\sin\theta~,\\
&\sigma^{\hat{\theta}}=-\sigma^3\sin\theta+(\sigma^1\cos\phi+\sigma^2\sin\phi)\cos\theta~,\\
&\sigma^{\hat{\phi}}=-\sigma^1\sin\phi+\sigma^2\cos\phi~. 
\end{split}\end{align}
If we now frame rotate by 
\eq{\label{framerot}
U^{~\beta}_{\alpha}=e^{\frac{i \phi}{2}\sigma^0\bar\sigma^3}e^{\frac{i\theta}{2}\sigma^0\bar\sigma^2}~,
}
then we find that the Dirac operator becomes 
\eq{
U^\dagger \bar\sigma^\mu D_\mu U=\sigma^3\left(\partial_r+\frac{1}{r}\right)+\frac{1}{r}\left(\sigma^1\partial_\theta+\frac{1 }{\sin(\theta)}\left(\sigma^2\partial_\phi+\frac{i \gamma_m\sigma^2}{2}-\half (i\sigma^2 \gamma_m+\sigma^1)\cos\theta\right)\right)~.
}

Now let us act with the Dirac operator on a fermion $\psi_{\mu_i}$ of charge $\langle \gamma_m,\mu_i\rangle=q_i$. 
Under the field redefinition
\be
\psi_{\mu_i}=\frac{e^{i\left( m-\frac{q_i}{2}\right)\phi}}{r} U\hat\psi_{\mu_i}~,
\ee
the Dirac operator acting on $\hat\psi_{\mu_i}$ further simplifies to 
\be
\bar\sigma^\mu D_\mu =\left(\sigma^3\partial_r-\frac{\CK }{r}\right)\quad, \quad \CK =\left(\begin{array}{cc}0&L^+_{m,\frac{q_i-1}{2}}\\
L^-_{m,\frac{q_i+1}{2}}&0\end{array}\right)^{\dot\alpha\alpha}~,
\ee
where 
\be
L^\pm_{m,m'}=\partial_\theta\pm \frac{1}{\sin\theta}(m-m'\cos\theta)~. 
\ee
The $L^\pm_{m,m'}$ operators are diagonalized by the Wigner small $d$ functions:
\be
L^\pm_{m,m'}d^j_{m,m'}(\theta)=\mp \sqrt{j(j+1)-m'(m'\pm1)}d^j_{m,m'\pm 1}(\theta)~, 
\ee
which are only defined for
\eq{
j\in \half \IZ_{\geq0}\quad, \quad j-m,j-m'\in \IZ\quad, \quad |m|\,,\,|m'|\leq j~.
}
We can then further reduce the ansatz for the Dirac operator
\eq{
\hat\psi_{\mu_i}=\left(\begin{array}{c}
f_+(r)\,	 d^j_{-m,\frac{q_i+1}{2}}(\theta)\\
f_-(r)\, d^j_{-m,\frac{q_i-1}{2}}(\theta)
\end{array}\right)~,
}
on which the Dirac operator acts as 
\eq{
\bar\sigma^\mu D_\mu \hat\psi_{\mu_i}=\left(\sigma^3\partial_r+\frac{i \sigma^2}{r} \sqrt{\left(j+\half\right)^2-\frac{q_i^2}{4}}\right)\hat\psi_{\mu_i}~. 
}
We can then solve for the spectrum of the Dirac operator by solving the matrix ODE. 
%\eq{
%i\bar\sigma^\mu D_\mu \psi_{\mu_i}=k\psi_{\mu_i}~,}
While the spectrum is somewhat involved the spectrum of $k=0$ modes are explicitly written 
\eq{
\psi_{\mu_i}^{(k,m)}(r,\theta,\phi)=&c_{+}\,\frac{e^{i\left(m- \frac{q_i}{2}\right)\phi}}{r} r^{\sqrt{(j+1/2)^2-q_i^2/4}}\,U(\theta,\phi)\left(\begin{array}{c}
 d^j_{-m,\frac{q_i+1}{2}}(\theta)\\
 d^j_{-m,\frac{q_i-1}{2}}(\theta)
\end{array}\right)\\
&+c_{-}\, \frac{e^{i\left(m- \frac{q_i}{2}\right)\phi}}{r} r^{-\sqrt{(j+1/2)^2-q_i^2/4}}\,U(\theta,\phi)\left(\begin{array}{c}
 d^j_{-m,\frac{q_i+1}{2}}(\theta)\\
- d^j_{-m,\frac{q_i-1}{2}}(\theta)
\end{array}\right)~.
}
Requiring that the energy density be finite at $r\to 0,\infty$ then implies that there are no solutions for $q_i=0$ and restricts us to the solutions for which 
\eq{
j=j_i:=\frac{|q_i|-1}{2}~,
}
in the case for $q_i\neq 0$. 
When restricted to this class of solutions, which are spin polarized, the full spectrum of the Dirac operator is written
\eq{
\psi_{\mu_i}^{(k,m)}=\frac{U(\theta,\phi) }{r}\hat\psi_{\mu_i}^{(k,m)}=\frac{U(\theta,\phi) }{r}\begin{cases}e^{i k r}e^{i(m-q_i/2)\phi}\left(\begin{array}{c}
d^{j_i}_{-m,j_i}(\theta)\\0
\end{array}\right)&q_i>0\\
e^{-i kr}e^{i(m-q_i/2)\phi}\left(\begin{array}{c}0\\
d^{j_i}_{-m,-j_i}(\theta)
\end{array}\right)&q_i<0
\end{cases}
}
We call these modes 
plane-wave normalizable because the $1/r$ behavior of $\psi$ is the leading term in the partial wave expansion. These solutions have time-dependence of the form 
\eq{
\psi_{\mu_i}^{(k,m)}=\frac{U(\theta,\phi) }{r}\hat\psi_{\mu_i}^{(k,m)}=\frac{U(\theta,\phi) }{r}\begin{cases}e^{i k(t+ r)}e^{i(m-q_i/2)\phi}\left(\begin{array}{c}
d^{j_i}_{-m,j_i}(\theta)\\0
\end{array}\right)&q_i>0\\
e^{i k(t-r)}e^{i(m-q_i/2)\phi}\left(\begin{array}{c}0\\
d^{j_i}_{-m,-j_i}(\theta)
\end{array}\right)&q_i<0
\end{cases}
}
Thus, we see that there are $|q_i|$ polarized, spherical fermion modes in the IR for each $q_i\neq 0$ that are in-going for $q_i>0$ and out-going for $q_i<0$. 

\subsection{The Abelian Monopole in $SU(N)$}

In order to compare the solutions of the Dirac equation in the abelian monopole background to the fermion zero-modes in the non-abelian monopole background, we need to solve for the fermion modes in the canonical gauge. This can be written as
\eq{
A_{can.}= T_3 (1-\cos\theta )d\phi\pm i \frac{I_\pm}{2}e^{\mp i \phi}(d\theta \mp i \sin\theta d\phi)~,
}
where here $e^{4\pi i T_3}=\mathds{1}_{SU(N)}$. After the frame rotation \eqref{framerot} the time independent Dirac operator can be written 
\eq{
\Bigg[\sigma^3\left(\partial_r+\frac{1}{r}\right)+\frac{\CK}{r}%\sigma^1\left(\partial_\theta+\half \cot(\theta)\right)+\frac{1}{r\sin\theta}\sigma^2&\left(\partial_\phi-\frac{i T_3}{2}-\frac{i \cos(\theta)}{2}T_3\right)\\
&-\left(\sigma^-e^{-i \phi}I_+-\sigma^+ e^{i \phi}I_-\right)\Bigg]~,
}
where
\eq{\label{ckdef}
\CK&=
\sigma^1\left(\partial_\theta+\half \cot(\theta)\right)+\frac{1}{\sin\theta}\sigma^2\Big((\partial_\phi+i T_3)-i \cos(\theta)T_3\Big)=\left(\begin{array}{cc}
0&D^+_{T_3-\half}\\
D^-_{T_3+\half}&0\end{array}\right)~,%^{\dot\alpha\alpha}
}
and
\eq{
D^\pm_{q}=\partial_\theta\pm \frac{1}{\sin\theta}\Big(-i(\partial_\phi+i T_3)-q\cos\theta\Big)~. 
}
Now let us consider the time-independent Dirac equation acting on a fermion $\psi_R$ of representation $R$ of $SU(N)$. Let us decompose $\psi_R$ into weights
\eq{
\psi_R=\sum_{\mu_a\in \Delta_R}\psi_R^a\,v_{\mu_a}~.
}
$\CK$ acts simply on the $SU(2)_J$-representations such that given a spin-$j$ representation:
\eq{
\psi_{R}^{a\,(j)} =U(\theta,\phi)e^{i(m-q_a/2)\phi}\left(\begin{array}{c}
f_{a,+}(r)\,d^{j}_{-m,\frac{q_{a}+1}{2}}(\theta)\\
f_{a,-}(r)\,d^{j}_{-m,\frac{q_{a}-1}{2}}(\theta)
\end{array}\right)\quad, \quad \mu_a\in \Delta^{(j)}_R~,
}
where $\langle T_3,\mu_a\rangle=\frac{q_a}{2}$ and $\psi_{R_A}^{(j)}$ is the spin-$j$ solution of $\psi_{R_A}$. Here, 
$\CK$ acts as
\eq{
\CK \cdot \psi_{R_A}^{a\,(j)}=\left(\begin{array}{cc}
0&-\sqrt{(j+1/2)^2-\frac{q_a^2}{4}}\\
\sqrt{(j+1/2)^2-\frac{q_a^2}{4}}&0
\end{array}\right)\psi_{R_A}^{a\,(j)}~.
}
Now, the Dirac equation reduces to a first order matrix ODE for the functions $f_{a,\pm}(r)$:
\eq{
\Big[r \partial_r+\CD\Big]F(r)=0~,
}
where 
\eq{
\CD=\left(\begin{array}{cc}
0&-\sqrt{(j+1/2)^2-\frac{q_a^2}{4}}\\
-\sqrt{(j+1/2)^2-\frac{q_a^2}{4}}&0
\end{array}\right)+\sigma^-I_++\sigma^+I_-+1~,
}
and 
\eq{
F(r)=\sum_{\mu_a\in \Delta_{R_A}^{(j)}}\left(\begin{array}{c}
f_{a,+}(r)\\
f_{a,-}(r)
\end{array}\right)\, v_{\mu_a}~.
}
The  solutions are given by 
\eq{
F(r)=P\,\exp\left\{-\int \CD(r)dr\right\}\,F_0~,
}
for $F_0$ a constant vector. The set of $F_0$ that corresponds to plane-wave normalizable solutions 
\eq{
\lim_{r\to \infty}r \psi^{(j)}_{R_A}<\infty~,
}
were solved for in \cite{Brennan:2021ucy}. The solutions decompose with respect to the decomposition of $\vI=\bigoplus_i \vI^{(i)}$. Let us denote the decomposition by the set of spins $\{s_I\}$ of the representations $\{\vI^{(i)}\}$ and $R_{s_I}$ to be the restriction of the representation $R$ to the spin-$s_I$ representation of $\vI^{(i)}$. By construction, for each $s_I$, the magnetic charge is a constant $Q_{s_I}$ on $\psi_R$. The $s_I$-solutions of  spin-$j$ only exist for $j_{min}^{(s_I)}\leq j\leq j_{max}^{(s_I)}$ where
\eq{
j_{min/max}^{(s_I)}=min/max_{\mu\in \Delta_{R_{s_I}}}|\langle \mu,T_3\rangle|-\half ~.
}
The solutions are given explicitly by 
\eq{
f_{a,+}(r)=0\quad, \quad f_{a,-}(r)=\frac{1}{r}\Big\langle j,Q_{s_I}-\half+s_I-a\,\Big{|}\,|Q_{s_I}|-\half,Q_{s_I}-\half;s_I,s_I-a\Big\rangle~,
}
for $Q_{s_I}>0$ and
\eq{
f_{a,+}(r)=\frac{1}{r}\Big\langle j,Q_{s_I}+\half+s_I-a\,\Big{|}\,
|Q_{s_I}|-\half,Q_{s_I}+\half;s_I,s_I-a\Big\rangle\quad, \quad f_{a,-}(r)=0~,
}
for $Q_{s_I}<0$  
 where 
 $\mu_a\in \Delta_{R_{s_I}}$ where we have chosen an ordering $\langle I_3,\mu_a\rangle>\langle I_3,\mu_{a+1}\rangle$. 
Note that there is no solution for modes with $Q_{s_I}=0$.  
 
   Importantly these modes are spin polarized. This means that the spectrum of the Dirac operator is then described by the solutions 
\eq{\label{canonicalasymptoticsolutions}
\psi_{R}^{a\,(j)} =\frac{e^{i (m-q_a/2)\phi}}{r}U(\theta,\phi)\begin{cases}e^{i k(t+r)}
\left(\begin{array}{c}
0\\
C^{(j)}_{s_I,a,-} \,d^{j}_{-m,\frac{q_{a}-1}{2}}(\theta)
\end{array}\right)&Q_{s_I}>0\\
e^{i k(t-r)}\left(\begin{array}{c}
C^{(j)}_{s_I,a,+}\, d^{j}_{-m,\frac{q_{a}+1}{2}}(\theta)\\
0
\end{array}\right)&Q_{s_I}<0
\end{cases}
}
for $\mu_a\in \Delta^{(j)}_{R_{s_I}}$ and 
\eq{
C^{(j)}_{s_I,a,\pm}=\Big\langle j,Q_{s_I}\pm \half+s_I-a\,\Big{|}\,|Q_{s_I}|-\half,Q_{s_I}\pm \half;s_I,s_I-a\Big\rangle~.
}
Note that in the cases where $\phi_\infty=0$, as we consider in the bulk of this paper, the theory has a preserved non-abelian gauge symmetry along the irreducible components $\vI^{(i)}$. This means that the above solutions are gauge equivalent to the solutions where $\psi^{(j)}_{R}$ is purely along a single weight $\mu\in \Delta_{R_{s_I}}$. In the bulk of the paper, we will find it convenient to pick a gauge such that each solution is along the top/bottom weight $\mu_{top/bot}\in \Delta_{R_{s_I}}$ where $|\langle T_3,\mu_{top/bot}\rangle|=j+\half$.

\subsection{The Non-Abelian Monopole}

Now let us solve for solutions to the Dirac operator coupled to the $SU(N)$ non-abelian monopole connection. Here we take the connection to be
\eq{
A=T_3(1-\cos\theta)d\phi+\ihalf M_+(r) e^{-i\phi}(d\theta-i \sin\theta d\phi)-\ihalf M_-(r)e^{i\phi}(d\theta+i \sin\theta d\phi)~.
}

Here we will be interested in solving for the zero-mode solutions which are spanned by the normalizable solutions in the kernel of the time-independent Dirac operator. Again, after performing a frame rotation,  the Dirac operator can be written 
\eq{
\Bigg[\sigma^3\left(\partial_r+\frac{1}{r}\right)+\frac{\CK}{r}%\sigma^1\left(\partial_\theta+\half \cot(\theta)\right)+\frac{1}{r\sin\theta}\sigma^2&\left(\partial_\phi-\frac{i T_3}{2}-\frac{i \cos(\theta)}{2}T_3\right)\\
&-\left(\sigma^-e^{-i \phi}M_+-\sigma^+ e^{i \phi}M_-\right)\Bigg]~,
}
where
\eq{\label{ckdef}
\CK&=
\sigma^1\left(\partial_\theta+\half \cot(\theta)\right)+\frac{1}{\sin\theta}\sigma^2\Big((\partial_\phi+i T_3)-i \cos(\theta)T_3\Big)=\left(\begin{array}{cc}
0&D^+_{T_3-\half}\\
D^-_{T_3+\half}&0\end{array}\right)~,%^{\dot\alpha\alpha}
}
and
\eq{
D^\pm_{q}=\partial_\theta\pm \frac{1}{\sin\theta}\Big(-i(\partial_\phi+i T_3)-q\cos\theta\Big)~. 
}

Now let us consider the time-independent Dirac equation acting on a fermion $\psi_R$ of representation $R$ of $SU(N)$. Let us decompose $\psi_R$ into weights
\eq{
\psi_R=\sum_{\mu_a\in \Delta_R}\psi_R^a\,v_{\mu_a}~.
}
$\CK$ acts simply on the $SU(2)_J$-representations such that given a spin-$j$ representation:
\eq{
\psi_{R_A}^{a\,(j)} =U(\theta,\phi)e^{i (m-q_a/2)\phi}\left(\begin{array}{c}
f_{a,+}(r)\, d^{j}_{-m,\frac{q_{a}+1}{2}}(\theta)\\
f_{a,-}(r)\,d^{j}_{-m,q_{a}-\half}(\theta)
\end{array}\right)\quad, \quad \mu_a\in \Delta^{(j)}_{R_A}~,
}
where $\langle T_3,\mu_a\rangle=\frac{q_a}{2}$ and $\psi_R^{(j)}$ is the spin-$j$ solution of $\psi_R$. Here, 
$\CK$ acts as
\eq{
\CK \cdot \psi_{R_A}^{a\,(j)}=\left(\begin{array}{cc}
0&-\sqrt{(j+1/2)^2-\frac{q_a^2}{4}}\\
\sqrt{(j+1/2)^2-\frac{q_a^2}{4}}&0
\end{array}\right)\psi_{R_A}^{a\,(j)}~.
}
Now, the Dirac equation reduces to a first order matrix ODE for the functions $f_{a,\pm}(r)$:
\eq{
\Big[r \partial_r+\CD\Big]F(r)=0\quad, \quad F(r)=\sum_{\mu_a\in \Delta_{R_A}^{(j)}}\left(\begin{array}{c}
f_{a,+}(r)\\
f_{a,-}(r)
\end{array}\right)\, v_{\mu_a}~,
}
where 
\eq{
\CD=\left(\begin{array}{cc}
0&-\sqrt{(j+1/2)^2-\frac{q_a^2}{4}}\\
-\sqrt{(j+1/2)^2-\frac{q_a^2}{4}}&0
\end{array}\right)+\sigma^-M_++\sigma^+M_-+1~.}
The smooth solutions given by 
\eq{
F(r)=P\,\exp\left\{-\int \CD(r)dr\right\}\,F_0~,
}
for $F_0$ a constant vector. As proven in \cite{Brennan:2021ucy}, there exists a vector $F_0$ that corresponds to a smooth, plane-wave normalizable solution 
\eq{
\lim_{r\to \infty}r \psi^{(j)}_{R_A}<\infty~,
}
of the time-independent Dirac equation iff the top and bottom weights of the spin-$j$ representation of $R_{A}$: 
  \eq{
\langle \mu_{top},T_3\rangle=-\langle \mu_{bot},T_3\rangle=j+\half\quad, \quad \mu_{top},\mu_{bot}\in \Delta^{(j)}_{R_A}~,
}
have charge of opposite sign under $\gamma_m$:
\eq{\label{topbotexist}
\langle \mu_{top},\gamma_m\rangle\langle \mu_{bot},\gamma_m\rangle<0~.
}
Note that since the asymptotic form of the connection approaches the abelian connection in the canonical gauge, the asymptotic form of the fermion zero-modes can be expressed as a linear combination of the zero-momentum solutions found in the previous subsection. 

As shown in \cite{Brennan:2021ucy}, for each irreducible $SU(2)_T$-representation $R_A$ and $j$ that satisfies \eqref{topbotexist} there is a unique solution which has the asymptotic form of 
\eq{
\lim_{r\to \infty}\psi_{R_A}^{(j)}=\psi_{R_A}^{(j)-}- \psi_{R_A}^{(+)}~, 
}
where $\psi_{R_A}^{(j)\pm}$ are given by the spin up/down solutions of \eqref{canonicalasymptoticsolutions}. %Here the sign is not fixed, but the two choices are equivalent up to a gauge transformation. 

\bibliographystyle{utphys}
\bibliography{MonopoleFermionScatteringBib}

\end{document}